\documentclass[number,review,3p]{elsarticle}
\usepackage[utf8]{inputenc}
\usepackage[T1]{fontenc}
\usepackage[british]{babel}
\usepackage{subcaption}
\usepackage[table,xcdraw]{xcolor}
\usepackage{textcomp}
\usepackage{multirow}
\usepackage[version=4]{mhchem}
\usepackage[most]{tcolorbox}
\usepackage[hyphens]{url}
\usepackage[hidelinks]{hyperref}
\hypersetup{breaklinks=true,
            pdfauthor=author}
\graphicspath{ {images_v3/} }
\usepackage{lineno}
\usepackage{soulutf8}
\usepackage{bm}
\usepackage{amsmath}
\usepackage{amssymb}
\usepackage{derivative}
\usepackage{todonotes}
\usepackage{siunitx}
    \DeclareSIUnit{\molarity}{molal}
\usepackage[font=small,labelfont=bf]{caption}
\usepackage{lipsum}
\usepackage{cleveref}
\usepackage{chemformula}
\usepackage[gen]{eurosym}
\usepackage{textcomp}
\usepackage{float}
\usepackage{pdflscape}
\usepackage{booktabs,tabularx,array}
\newcolumntype{L}[1]{>{\raggedright\arraybackslash}p{#1}}
\newcolumntype{C}[1]{>{\centering\arraybackslash}p{#1}}

\makeatletter
\def\ps@pprintTitle{%
  \let\@oddhead\@empty
  \let\@evenhead\@empty
  \let\@oddfoot\@empty
  \let\@evenfoot\@oddfoot
}
\makeatother

\usepackage[tight-spacing=true]{siunitx} 
\usepackage{makecell} 
\usepackage{multicol}
\usepackage{nomencl}
\makenomenclature

\renewcommand\nompreamble{\begin{multicols}{2}} 
	\renewcommand\nompostamble{\end{multicols}}

\renewcommand\nomgroup[1]{%
	\item[\bfseries
	\ifstrequal{#1}{A}{Roman Letters}{}%
	\ifstrequal{#1}{B}{Greek Letters}{}%
	\ifstrequal{#1}{C}{Subscripts and superscripts}{}%
	\ifstrequal{#1}{D}{Abbreviations}{}%
	\ifstrequal{#1}{E}{Superscripts}{}%
	]}
\usepackage{svg}

\makeatletter
\renewcommand\@makefntext[1]{\leftskip=2em\hskip-2em\@makefnmark#1}
\makeatother

\begin{document}
\begin{frontmatter}
    \title{\textcolor{black}{When can buoyancy be neglected in a confluence flow cell? Evidence from crystallization fouling experiments}}
\author[1,2]{Isaac Appelquist Løge\corref{cor1}}
\ead{isaac.a.loge@ntnu.no}

\author[1,5]{Candela Almazán Pérez}
\ead{candela.almazanperez@gmail.com}

\author[3,4]{Maike W. Baltussen}
\ead{m.w.baltussen@tue.nl}

\author[1]{Nadia Shardt}
\ead{nadia.shardt@ntnu.no}

\cortext[cor1]{Corresponding author}

\affiliation[1]{
    organization={Department of Chemical Engineering,
                  Norwegian University of Science and Technology (NTNU)},
    city={Trondheim},
    postcode={7046},
    country={Norway}
}

\affiliation[2]{
    organization={Technical University of Denmark,
                  Department of Chemical Engineering},
    addressline={Søltofts Plads 228A},
    city={Kgs. Lyngby},
    postcode={2800},
    country={Denmark}
}
\affiliation[5]{
    organization={Department of Chemical Engineering, Universidad Autónoma de Madrid (UAM)},
    addressline={Faculty of Sciences, C/ Francisco Tomás y Valiente, 7, Campus de Cantoblanco},
    city={Madrid},
    postcode={28049},
    country={Spain}
}

\affiliation[3]{
    organization={Multiphase Flows for Energy Applications,
                  Department of Chemical Engineering and Chemistry,
                  Eindhoven University of Technology},
    addressline={P.O. Box 513},
    city={Eindhoven},
    postcode={5600 MB},
    country={The Netherlands}
}
\affiliation[4]{
    organization={Eindhoven Institute for Renewable Energy Systems, Eindhoven University of Technology},
    addressline={P.O. Box 513},
    city={Eindhoven},
    postcode={5600 MB},
    country={The Netherlands}
}

    \begin{abstract}
        \textcolor{black}{Predicting crystallization fouling remains difficult because fouling combines several coupled processes: supersaturation, nucleation, crystal growth, wall transport, and adhesion. Separating these processes requires model systems in which each can be controlled. One such system is the confluence flow cell, in which two incompatible brines meet and mix, generating supersaturation at the interface. Where the streams meet is usually predicted with simulations that assume constant density, because the density difference between the brines is only a few percent or less.} 

In this paper, we investigate whether \textcolor{black}{assuming constant density} holds in a \textcolor{black}{confluence flow cell} where calcium chloride and sodium sulfate meet and form gypsum. A constant density computational fluid dynamics simulation of the reacting flow results in deposition on the floor along the channel centreline. However, in the experiment, gypsum instead forms bands on both side walls, and the deposition bands start further upstream as the flow rate decreases. Dye experiments with density differences between the streams of 0.14 to 2.1\,\%, three channel heights, and 32 flow rates show that the distance at which the dyed stream reaches the opposite side wall is governed by the bulk Richardson number, $Ri=g'H/U^2$, and follows $L/H = 3.31\,Ri^{-0.66}$ over four decades of $Ri$ ($R^2=0.89$). When the two streams' densities match, the dyed stream no longer reaches the opposite wall under most conditions, and in a density-matched reacting experiment, gypsum deposits on the floor, as the constant-density simulation predicts. 

A small relative density difference is thus not a sufficient condition for neglecting buoyancy: the Richardson number must also be small. \textcolor{black}{At the low flow rates typical of such flow cells}, it often is not. \textcolor{black}{Because $Ri$ depends only on the density difference, channel height, and flow velocity, we recommend evaluating it before using a constant-density model for mixing, reaction, or fouling.}
    \end{abstract}
\end{frontmatter}


\section{Introduction}
\label{sec:intro}
Confluence channels bring two separately supplied liquid streams into contact under controlled conditions. Because \textcolor{black}{of the control it allows over inlet compositions and flow rate,} researchers widely use them to study mixing, reactions, and the spatial distribution of dissolved species~\cite{yoon2005}. \textcolor{black}{One such application is crystallization fouling, the deposition of sparingly soluble salts on heat-transfer and process surfaces, which remains a persistent and costly problem in heat-exchanger operation~\cite{mullersteinhagen2011}.} \textcolor{black}{Predicting the magnitude of fouling is not sufficient}, as the operational consequences of fouling also depend on its spatial distribution\textcolor{black}{, and numerical models of fouling increasingly resolve where on a surface deposits form, grow and detach~\cite{brahim2003,loge2022,loge2024}}. \textcolor{black}{Where deposits form in a flow cell} depends on both how dissolved precursors are transported to the walls and the thermodynamic driving force for precipitation.

The usual way to model such a cell is to compute the velocity field assuming constant fluid properties, then solve the advection-diffusion equations for the dissolved species on that field, without \textcolor{black}{coupling the composition back to the flow}. Sato et al.\ adopt the same density and viscosity for the sample and sheath streams in a Y-shaped microchannel and treat dilute ions and particles as passive scalars~\cite{sato2021}; Okuducu and Aral likewise assume constant density and viscosity and neglect gravity when computing scalar transport in a passive micromixer \cite{okuducu2021}. The justification is that the solutions are dilute, and the density difference between the streams is a small fraction of the density. \textcolor{black}{This assumption is often adequate in microchannels, but the small density difference alone does not guarantee it}, and it is not obvious that it carries over to a larger, more slowly operated precipitation cell.

A small difference in relative density justifies simplifying the equations of motion but not removing buoyancy from them. In the Boussinesq approximation, the density difference is neglected in the inertia of the fluid but retained in the gravitational force. The importance of the gravitational force can only be determined by comparing it with the inertia of the imposed flow. \textcolor{black}{For a density difference $\Delta\rho$ between the streams, a reference density $\rho_0$, a channel height $H$, and a bulk velocity $U$, the reduced gravity, the buoyancy velocity, and the bulk Richardson number are~\cite{turner1973,benjamin1968}:}

\begin{equation}
    g'=g\frac{\Delta\rho}{\rho_0},
    \qquad U_b=\sqrt{g'H},
    \qquad Ri=\frac{g'H}{U^2}=\left(\frac{U_b}{U}\right)^2 .
    \label{eq:intro_bulk_ri}
\end{equation}

\textcolor{black}{A small value of $\Delta\rho/\rho_0$ does not imply a small value of $Ri$, because $Ri$ also scales with the channel height and the inverse square of the velocity. In a confined, viscous device, $U_b$ indicates only the order of magnitude of buoyancy-driven velocities.}

\textcolor{black}{Buoyancy has previously been shown to reorganize the flow in small confluence channels.} Yoon et al.\ showed that the interface between two liquids of different density flowing side by side in a microchannel reorients under gravity~\cite{yoon2005}. Mariotti et al.\ combined micro-PIV, flow visualization, and direct numerical simulation to show how stratification from a small density difference affects mixing and reaction yield in a T-shaped micro-mixer, and described the stratified regime with Richardson-number scaling~\cite{mariotti2021}. 

However, confinement makes the transfer conditional. The top and bottom walls, side-wall drag, the cross-sectional aspect ratio, and finite residence time all affect how two streams of different density redistribute. Classical gravity-current theory already treats bounded configurations~\cite{benjamin1968}, and experiments on viscous lock-exchange flows \textcolor{black}{(in which two fluids of different density, initially separated by a vertical gate, are released and flow under and over each other)} in rectangular channels show that the balance between gravitational driving and viscous resistance depends on geometry~\cite{martin2011}. In microfluidic interdiffusion, Salmon et al.\ identify the solutal convection regimes using a Rayleigh number and elapsed time~\cite{salmon2021}, and in Hele-Shaw geometries, convective instability of diffusion layers controls mass transport~\cite{backhaus2011}. \textcolor{black}{Therefore, we must test a correlation in terms of a bulk Richardson number in the device of interest, alongside the Reynolds number, molecular diffusion, and geometry}: reducing $U$ increases $Ri$, but it does not remove the viscous constraints on the motion.

This paper examines that question in a stepped confluence flow cell operating with density differences of 0.14 to 2.1\,\% between streams. We first set up a reacting flow simulation that neglects these density differences and compare its predicted gypsum deposition locations with experimental observations. Based on dye experiments with varying density differences, channel height, and flow rate, we show that the dense stream's transport is governed by the bulk Richardson number. The contribution is not a new criterion for buoyancy but a controlled demonstration, in a practical fouling geometry, that the small-density-difference assumption fails when the Richardson number is not \textcolor{black}{small as well}. \textcolor{black}{From the clear-zone correlation obtained in this cell, we also give a practical threshold below which buoyancy can be neglected: the buoyancy-driven spreading has no effect within a device of length $L_{\mathrm{dev}}$ when $Ri<(3.31\,H/L_{\mathrm{dev}})^{1/0.66}$, which for the present $120\ \mathrm{mm}$ cell gives $Ri<0.065$, $0.11$ and $0.19$ for $H=6.0$, $8.5$ and $12.3\ \mathrm{mm}$.}

\section{Experimental Materials and Method}
\subsection{Flow cell}
\label{sec:cell}
\begin{figure*}[htbp]
    \centering
    \includegraphics[width=\linewidth]{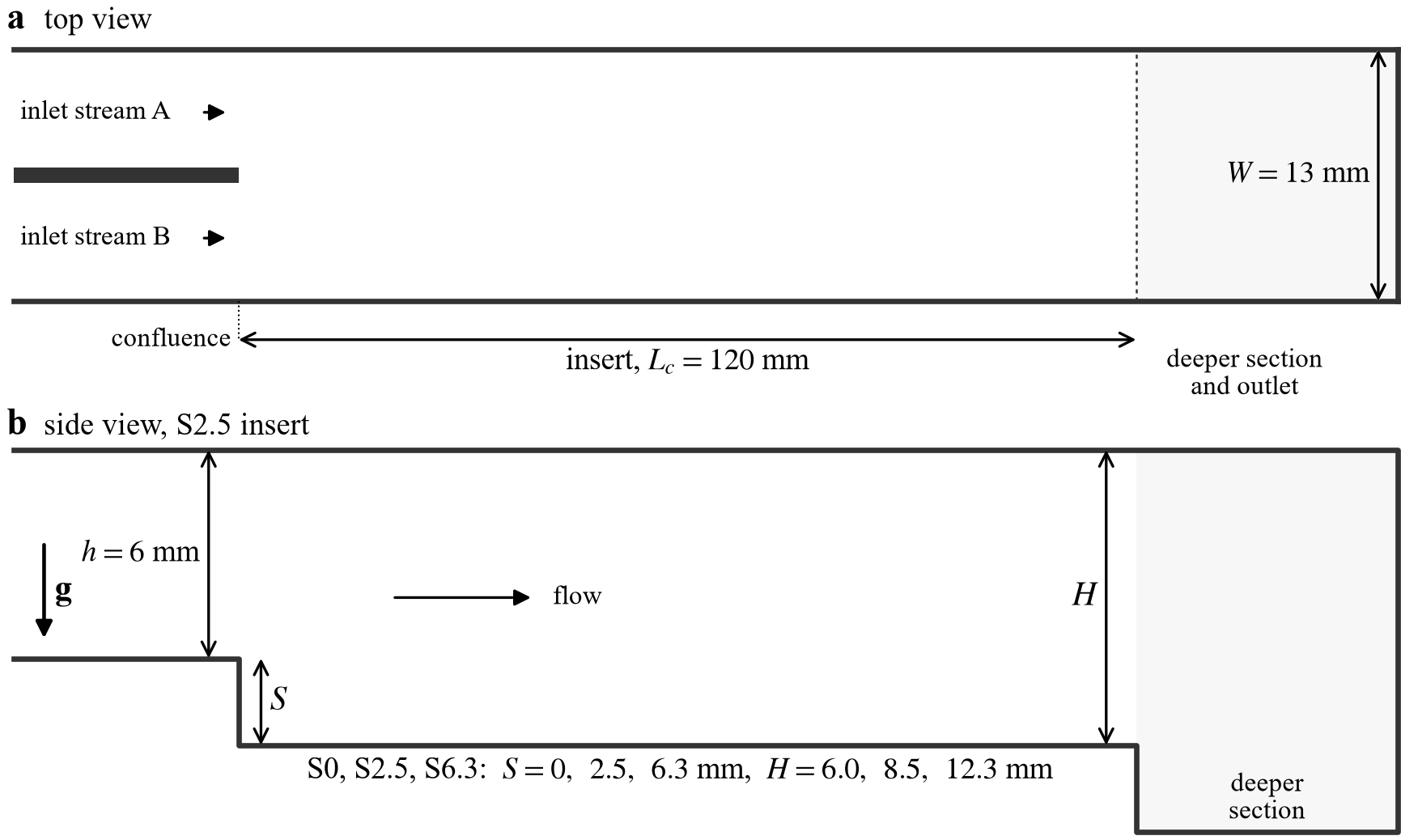}
    \caption{The flow cell (schematic; heights exaggerated).
    \textbf{a} Top view. The two streams meet at the confluence and flow through the $120\ \mathrm{mm}$ insert. \textbf{b} Side view of the \textcolor{black}{S2.5} insert.}
    \label{fig:cell}
\end{figure*}

The experiments were performed in a rectangular flow cell \textcolor{black}{(\cref{fig:cell}) with a steel body and a PMMA lid} of width $W=13.0\ \mathrm{mm}$ and upstream channel height $h=6.0\ \mathrm{mm}$. Two streams enter through separate inlet channels and meet at a confluence. At the confluence, the channels have an exchangeable insert, which can change the height of the channel. The channel insert has a length of $L_c=120\ \mathrm{mm}$, after which the channel expands to a height of 11\,mm over 35\,mm, terminating in a cylindrical outlet. Three interchangeable inserts impose a backward-facing step at the confluence at heights $S=0$, $2.5$ and $6.3\ \mathrm{mm}$ (denoted \textcolor{black}{S0, S2.5 and S6.3}, respectively), giving downstream channel heights $H=6.0$, $8.5$ and $12.3\ \mathrm{mm}$. The bulk velocity downstream of the step is $U=2Q/(WH)$, where $Q$ is the volumetric flow rate \textcolor{black}{of each of the two inlets; the factor 2 accounts for both streams passing through the cross-section}.

Flow was delivered by two peristaltic pumps\textcolor{black}{, one per inlet (Kamoer FX-STP~3, closed-loop stepper motor with a three-roller KCS3 pump head and $3.2\times6.4\ \mathrm{mm}$ tubing)}. The pumps were calibrated \textcolor{black}{against} the measured pump frequency\textcolor{black}{: each pump cycle delivers $162\ \mathrm{\mu L}$} per inlet ($0.10315\ \mathrm{Hz}$ per $\mathrm{mL\,min^{-1}}$ per inlet).

\subsection{Working fluids and measured density differences}
\label{sec:fluids}
All solutions were prepared by mass in deionized water containing about $0.1\ \mathrm{g}$ KOH per kilogram of water. \textcolor{black}{Anhydrous CaCl$_2$ (Sigma-Aldrich C1016, $\geq93.0\,\%$), Na$_2$SO$_4$ (Sigma-Aldrich 238597, ACS reagent, $\geq99\,\%$), NaCl, KOH and carmine were used as received.} One stream was dyed with carmine at a concentration of about $0.05\ \mathrm{g\,kg^{-1}}$. Densities were measured at $20.00\pm0.01\ ^{\circ}\mathrm{C}$ with an Anton Paar DMA 5000M.

The density differences studied here are small. In the dye experiments, NaCl was added to the undyed stream \textcolor{black}{until the density difference between the streams equaled that of} the reacting CaCl$_2$/Na$_2$SO$_4$ pair at a mixed concentration of $0.01$, $0.1$ or $0.3\ \mathrm{M}$. The dyed stream is therefore the \textcolor{black}{less dense one}. The resulting \textcolor{black}{relative density differences}, $\Delta\rho/\rho_0=0.14$ to $2.1\,\%$ (\cref{tab:density}), \textcolor{black}{are of the size that could be regarded as negligible} when modeling the mixing of aqueous streams\textcolor{black}{~\cite{sato2021,okuducu2021}}, and this was the basis for the simulations of \cref{sec:governing}. We labelled the three series by their \textcolor{black}{density difference}; the NaCl added was about $0.03$, $0.18$ and $0.5\ \mathrm{mol\,kg^{-1}}$. A fourth series used dyed and undyed water without salt, for which the measured difference was $0.04$ to $0.06\ \mathrm{kg\,m^{-3}}$.

\begin{table}[htbp]
\centering
\caption{Measured density differences between the two streams. The first three rows are the dye experiments (NaCl added to the undyed stream); the last three are the reacting salts, CaCl$_2$ against Na$_2$SO$_4$, whose \textcolor{black}{density difference} the dye experiments reproduce. Feed molalities are those of each stream before mixing.}
\label{tab:density}
\begin{tabular}{llcc}
\toprule
System & Series & $\Delta\rho\ [\mathrm{kg\,m^{-3}}]$ & $\Delta\rho/\rho_0\ [\%]$ \\
\midrule
Dye, NaCl in undyed stream & low    & $1.36 \pm 0.04$  & $0.136$ \\
                           & medium & $7.09 \pm 0.01$  & $0.710$ \\
                           & high   & $20.73 \pm 0.15$ & $2.077$ \\
\midrule
CaCl$_2$ / Na$_2$SO$_4$ & $0.020$ / $0.020\ \mathrm{mol\,kg^{-1}}$ & $1.01$  & $0.10$ \\
                        & $0.199$ / $0.201\ \mathrm{mol\,kg^{-1}}$ & $7.37$  & $0.74$ \\
                        & $0.593$ / $0.588\ \mathrm{mol\,kg^{-1}}$ & $20.04$ & $2.00$ \\
\bottomrule
\end{tabular}
\end{table}

\paragraph{Density-matched controls}
A separate set of dye experiments used the reacting salts themselves: CaCl$_2$ in one stream and Na$_2$SO$_4$ with carmine in the other, at mixed concentrations of $0.1$ and $0.3\ \mathrm{M}$, so that the dyed stream is the denser \textcolor{black}{one}. In the density-matched series, NaCl was added stepwise to the CaCl$_2$ stream until the two streams were matched in density. The remaining differences, $\rho_A-\rho_B$ as in \cref{tab:matched_reacting}, were $-0.19$ and $+0.16\ \mathrm{kg\,m^{-3}}$, i.e.,\ $2.6$ and $0.8\,\%$ of the unmatched \textcolor{black}{density differences}. Matching in this way keeps the dye, the reacting salts and the flow rates unchanged while removing almost all of the density difference. 

\subsection{Optical measurement of the clear-zone length}
\label{sec:optics}
The cell was filmed from above, so each image integrates light attenuation through the full channel height. For each series, the flow rate was increased from $5$ to $160\ \mathrm{mL\,min^{-1}}$ per inlet in increments of $5\ \mathrm{mL\,min^{-1}}$, after the lines had been purged of air, and each of the 32 flow rates was recorded for $30\ \mathrm{s}$. The dye signal was taken as $F=\log\!\left[R/\tfrac12(G+B)\right]$\textcolor{black}{, where $R$, $G$ and $B$ are the time-averaged intensities of the red, green and blue channels of the camera image,} and normalized between the clear and dyed levels of each video (15th and 85th percentiles). 

The \textcolor{black}{measured quantity} is the \emph{length of the clear zone} $L$: the distance downstream of the confluence at which the normalized signal, averaged over a strip adjacent to the side wall opposite the dyed inlet, \textcolor{black}{first exceeds $25\,\%$ of the way from the clear to the dyed level} and remains above it over five pixels. $L$ therefore marks \textcolor{black}{the position at which dye first becomes detectable, in the top view, next to the opposite wall}. Conceptually, the length of the clear zone is the same as the inversion width, as described by Yoon et al.~\cite{yoon2005}. It does not, by itself, show that dye has reached the floor or that the cross-section is mixed. Conditions \textcolor{black}{in which the dye did not reach the opposite wall} within the $120\ \mathrm{mm}$ channel \textcolor{black}{are recorded as $L=120\ \mathrm{mm}$, which is only a lower bound on $L$, and are} excluded from fits.

\subsection{Deposition experiments}
\label{sec:deposition}
To see where gypsum forms in the cell, CaCl$_2$ and Na$_2$SO$_4$ solutions at a mixed concentration of $0.312\ \mathrm{M}$ were fed through the two inlets of the S0 configuration, whose floor was an aluminium coupon. Runs were made at $16.2$, $32.3$ and $64.8\ \mathrm{mL\,min^{-1}}$ per pump and lasted between $25$ and $58\ \mathrm{min}$. Because the deposit formed on the side walls rather than on the floor, the two side walls and the floor were photographed after each run and combined into one unfolded view of the channel. The deposit was segmented in this view, and its position on each side wall was traced along the channel as the centreline.

In a final set of reacting experiments, the two feeds were density-matched with NaCl, as for the dye controls. These were run at a mixed concentration of $0.3\ \mathrm{M}$ and $20$, $50$ and $100\ \mathrm{mL\,min^{-1}}$ per inlet, with residual density differences of \textcolor{black}{$0.02$ to $0.15\ \mathrm{kg\,m^{-3}}$ in five of the seven feed pairs, and $0.59$ and $1.06\ \mathrm{kg\,m^{-3}}$ in the repeat at $20$ and the first run at $100\ \mathrm{mL\,min^{-1}}$.} \textcolor{black}{The individual values and the corresponding Richardson numbers are listed in \cref{tab:matched_reacting}; all runs except the $20\ \mathrm{mL\,min^{-1}}$ repeat ($Ri\approx0.5$ to $4$, depending on the insert) lie below the threshold $Ri\approx0.065$ for $H=6\ \mathrm{mm}$ derived in \cref{sec:collapse}.}
\begin{table}[htbp]
\centering
\caption{\textcolor{black}{Residual density differences of the density-matched reacting runs at a mixed concentration of $0.3\ \mathrm{M}$. $\Delta\rho=\rho_A-\rho_B$ is the density of the CaCl$_2$+NaCl feed minus that of the Na$_2$SO$_4$ feed (which carries the carmine), both measured at $20.00\ ^{\circ}\mathrm{C}$ after the last addition of NaCl; a positive value means that the CaCl$_2$ feed is the denser one. Each feed pair was prepared separately; the 26 June pair was run in all three inserts. $U=2Q/(WH)$ and $Ri=g|\Delta\rho|H/(\rho_0U^2)$ with $\rho_0=998.2\ \mathrm{kg\,m^{-3}}$.}}
\label{tab:matched_reacting}
\textcolor{black}{%
\begin{tabular}{lllccc}
\toprule
$Q$ per inlet [mL\,min$^{-1}$] & Date (2026) & Insert & $\Delta\rho$ [kg\,m$^{-3}$] & $U$ [mm\,s$^{-1}$] & $Ri$ [--] \\
\midrule
20             & 23 June & S0   & $+0.022$ & 8.5  & 0.018 \\
20 (repeat)    & 26 June & S0   & $-0.585$ & 8.5  & 0.47  \\
               &         & S2.5 &          & 6.0  & 1.3   \\
               &         & S6.3 &          & 4.2  & 4.1   \\
\addlinespace
50             & 24 June & S0   & $-0.123$ & 21.4 & 0.016 \\
50             & 24 June & S0   & $+0.055$ & 21.4 & 0.007 \\
50             & 24 June & S0   & $-0.098$ & 21.4 & 0.013 \\
\addlinespace
100            & 25 June & S0   & $-1.059$ & 42.7 & 0.034 \\
100 (repeat)   & --      & S0   & $-0.149$ & 42.7 & 0.005 \\
\bottomrule
\end{tabular}}
\end{table}

\subsection{\textcolor{black}{Image analysis}}
\label{sec:image_analysis}
\textcolor{black}{The recordings and photographs were prepared with in-house MATLAB scripts.}

\textcolor{black}{\textit{Dye recordings.} The recordings, taken at $100$ frames per second, were first brought to a common geometry. In a reference recording, the channel axis was marked and the frame rotated to align it with the image axes, the insert was cropped, and the confluence and a mask of the channel were marked. Every recording was aligned to this reference in the same way, so that all recordings share the crop and have their origin at the confluence. From each recording a window in which the flow was steady was selected from a probe signal and cut to the same length for all flow rates (about $20\ \mathrm{s}$). The image scale follows from the $120\ \mathrm{mm}$ between the confluence and the end of the insert and is $0.18$ to $0.31\ \mathrm{mm}$ per pixel. The dye signal, its normalization and the clear-zone length $L$ are defined in \cref{sec:optics}.}

\textit{Apparent front speed.} In each pixel column of the time-averaged, normalized image, the interface was located at the transverse position where the signal, counted from the wall opposite the dyed inlet, first exceeds $25\,\%$, and the resulting line was smoothed along the channel. Its distance from that wall, $y_\mathrm{i}(x)$, is the local width of the clear zone; it decreases from $y_\mathrm{i}(0)$ at the confluence to zero where the dye reaches the wall ($x=L$), or to $y_\mathrm{i}(L_c)$ at the end of the insert if it does not. With the advective time $t=x/U$, the apparent front speed is the mean rate at which the clear zone narrows,

\begin{equation}
    v_{\mathrm{app}}=\frac{y_\mathrm{i}(0)-y_\mathrm{i}(x_\mathrm{e})}{x_\mathrm{e}/U},
    \qquad x_\mathrm{e}=\min(L,\,L_c).
    \label{eq:vapp}
\end{equation}

It is therefore a projected, time-averaged rate of lateral spreading, not the speed of a front followed in time.

\textit{Deposition photographs.} Each face of the channel (two side walls and the floor) was rectified to its known dimensions by a projective transformation fitted to four manually selected corners, corrected for uneven illumination by dividing by a Gaussian-blurred copy of the image, and resampled to $5\ \mathrm{px\,mm^{-1}}$. The three faces were then joined into one unfolded view. Inside a hand-drawn search mask slightly larger than the deposit, each pixel was scored by its colour distance from the local background in CIELAB space, with the background taken as the median colour of a ring along the edge of the mask. The score was thresholded at Otsu's level shifted by a manually set offset, objects smaller than $20$ pixels were removed, and nearby fragments were joined by morphological closing. The band centreline on each side wall was taken from this mask (Supplementary Material, Section~S3).

\section{Simulation Methods}
\subsection{Governing equations and turbulence modelling}
\label{sec:governing}
All simulations were performed with the incompressible solver \texttt{incompressibleFluid} of OpenFOAM version~13 \cite{greenshields2025,openfoam13release}. For a Newtonian liquid whose density varies only slightly with composition, the Boussinesq form of the continuity and momentum equations is:

\begin{align}
\nabla\cdot \mathbf{U} &= 0,\\
\rho_0\left[\frac{\partial \mathbf{U}}{\partial t} + \nabla\cdot(\mathbf{U}\mathbf{U})\right]
&= -\nabla P + \nabla\cdot\left[\rho_0\left(\nu+\nu_t\right)
\left(\nabla\mathbf{U}+(\nabla\mathbf{U})^{T}\right)\right] + (\rho-\rho_0)\,\mathbf{g},
\label{eq:momentum}
\end{align}

where $\mathbf{U}$ is the mean velocity, $\rho_0$ a reference density, $P$ the pressure relative to hydrostatic, $\nu$ the kinematic viscosity and $\nu_t$ the eddy viscosity of the $k$--$\varepsilon$ model described below.

\paragraph{Neglecting buoyancy}
In \cref{eq:momentum}, composition acts on the flow only through the last term. The Boussinesq approximation requires $\Delta\rho/\rho_0\ll1$, which holds here with $\Delta\rho/\rho_0\le2.1\,\%$. Dropping the buoyancy term altogether requires \textcolor{black}{a second condition}: the buoyancy force must also be small compared with the inertia of the imposed flow, that is, the Richardson number $Ri=g'H/U^2$ of \cref{sec:dimensionless} must satisfy $Ri\ll1$. In the simulations reported here, the density was set to $\rho_0$ throughout, and no gravitational acceleration was defined, so the last term of \cref{eq:momentum} vanishes, and the transported species do not act back on the flow. This choice was based on the small value of $\Delta\rho/\rho_0$; the second condition was not checked when the simulations were set up and is evaluated in \cref{sec:discussion}. The equations solved are therefore \cref{eq:momentum} without the buoyancy term, divided by $\rho_0$, with the kinematic pressure $p=P/\rho_0$ and constant $\nu=8.9\times10^{-7}\ \mathrm{m^{2}\,s^{-1}}$ \textcolor{black}{(water at 25\,$^{\circ}$C~\cite{kestin1978})}.

\paragraph{Turbulence closure}
The eddy viscosity was obtained from the standard $k$--$\varepsilon$ model with the OpenFOAM default coefficients:

\begin{align}
\frac{\partial k}{\partial t} + \nabla\cdot(\mathbf{U}k)
&= \nabla\cdot\left[\left(\nu + \frac{\nu_t}{\sigma_k}\right)\nabla k\right]
+ P_k - \varepsilon, \\
\frac{\partial \varepsilon}{\partial t} + \nabla\cdot(\mathbf{U}\varepsilon)
&= \nabla\cdot\left[\left(\nu + \frac{\nu_t}{\sigma_\varepsilon}\right)\nabla \varepsilon\right]
+ C_1\frac{\varepsilon}{k}P_k - C_2\frac{\varepsilon^2}{k},
\end{align}

with $\nu_t = C_\mu k^2/\varepsilon$. 

A RANS closure was used because the quantity of interest is the time-averaged transport of the reactants to the walls, and hence the location of deposition, rather than the unsteady flow structure. Downstream of the step, the flow is laminar to transitional throughout the campaign (\textcolor{black}{$\mathrm{Re}_{D}\approx50$ to $1300$} in the inlet channel), a range in which the $k$--$\varepsilon$ model is not guaranteed to be accurate. The closure was therefore compared with scale-resolving simulations of the same cell at constant density, with a passive tracer in place of the reactants (Supplementary Material, Section~S1). 

First, large-eddy simulations with the WALE subgrid model at the experimental flow rates and with steady inlets remain steady at $40$ and $60\ \mathrm{mL\,min^{-1}}$ per inlet, which bracket the reference case ($48.5\ \mathrm{mL\,min^{-1}}$), and become unsteady between $60$ and $90\ \mathrm{mL\,min^{-1}}$. At the reference condition, the constant-density flow is therefore steady, and a time-averaged closure \textcolor{black}{does not discard any resolved unsteadiness}. 

Second, at higher flow rates the closure does matter: at $115\ \mathrm{mL\,min^{-1}}$ per inlet, $k$--$\varepsilon$ keeps a sharp, straight interface between the streams, whereas a delayed detached-eddy simulation based on the Spalart--Allmaras model (DDES-SA) and WALE LES both give a meandering interface and much faster lateral mixing. 

Third, in all these constant-density computations, including the LES, the interface stays vertical, spanning the channel height; the tracer distributions in horizontal planes near the floor, at mid-height, and near the lid are alike. The choice of closure therefore affects how fast the streams mix, but not the orientation of the interface. 

\subsection{Reactive transport, precipitation and wall deposition}
\label{sec:reactive}
To represent the mass transport in the channel\textcolor{black}{, three scalar fields were included:} \texttt{A}, \texttt{B} and \texttt{C}. Here \texttt{A} represents dissolved Ca$^{2+}$ molality, \texttt{B} dissolved SO$_4^{2-}$ molality, and \texttt{C} a bookkeeping field for precipitated CaSO$_4$, all \textcolor{black}{expressed as molalities, in moles per kilogram of water} ($\mathrm{mol\,kg_w^{-1}}$\textcolor{black}{, the unit of the thermodynamic lookup table}). Scalar transport was handled by OpenFOAM function objects executed each time step, while precipitation and deposition were applied as operator-split source updates through a custom function object.

Transport of the dissolved species includes molecular diffusion with constant diffusivities:

\begin{align}
\frac{\partial A}{\partial t} + \nabla\cdot(\mathbf{U}A) &= \nabla\cdot\!\left(D_A\nabla A\right) + S_A, \\
\frac{\partial B}{\partial t} + \nabla\cdot(\mathbf{U}B) &= \nabla\cdot\!\left(D_B\nabla B\right) + S_B,
\end{align}

with $D_A=7.93\times10^{-10}\ \mathrm{m^2\,s^{-1}}$ and
$D_B=1.07\times10^{-9}\ \mathrm{m^2\,s^{-1}}$\textcolor{black}{, the tracer diffusivities of Ca$^{2+}$ and SO$_4^{2-}$ at infinite dilution and 25\,$^{\circ}$C~\cite{li1974}}. The precipitated scalar \texttt{C} was
transported without molecular diffusion,

\begin{equation}
\frac{\partial C}{\partial t} + \nabla\cdot(\mathbf{U}C) = S_C .
\end{equation}

The source terms couple the three fields through precipitation. Each mole of CaSO$_4$ formed removes one mole of each ion, so $S_A=S_B=-(r_{\mathrm{bulk}}+r_{\mathrm{wall}})$, where $r_{\mathrm{bulk}}$ is the rate of precipitation in the liquid (\cref{eq:bulkPrecipRate}) and $r_{\mathrm{wall}}$ the rate of deposition in cells adjacent to a wall (\cref{eq:wallsink}). Only the bulk precipitate remains in suspension (i.e.,  $S_C=r_{\mathrm{bulk}}$), while the deposited amount accumulates on the wall faces. 

Precipitation was modelled as an irreversible conversion, $A+B\rightarrow C$, which occurs only under supersaturation. The local saturation ratio $\mathrm{SR}=\mathrm{SR}(A,B)$ was obtained by bilinear interpolation from a $100\times100$ thermodynamic lookup table calculated with PHREEQC\textcolor{black}{~\cite{parkhurst2013} and its Pitzer database (\texttt{pitzer.dat}) for CaSO$_4$ in water at 25\,$^{\circ}$C, without the Na$^{+}$ and Cl$^{-}$ of the feeds, which overestimates SR in the mixed feeds by roughly a third}, spanning $10^{-6}$ to $2.0\ \mathrm{mol\,kg_w^{-1}}$ in each ion. \textcolor{black}{For gypsum, the relative supersaturation was defined from the square root of the saturation ratio}:

\begin{equation}
\sigma = \max\!\left(\mathrm{SR}^{1/2} - X_{\mathrm{crit}}^{1/2},\,0\right),
\qquad X_{\mathrm{crit}}=1.0,
\label{eq:sigma}
\end{equation}

\textcolor{black}{where $X_{\mathrm{crit}}=1.0$ means that precipitation starts as soon as the solution is supersaturated,} and bulk precipitation \textcolor{black}{was} expressed as:

\begin{equation}
\frac{\mathrm{d}C}{\mathrm{d}t} = k_{\mathrm{bulk}}\,\sigma^{\,n_{\mathrm{bulk}}},
\label{eq:bulkPrecipRate}
\end{equation}

with $k_{\mathrm{bulk}}=3.8\times10^{-6}\ \mathrm{mol\,kg_w^{-1}\,s^{-1}}$ and $n_{\mathrm{bulk}}=2.0$. \textcolor{black}{This form of the driving force, with two ions per formula unit, is the standard rate law for gypsum growth, which is of second order close to equilibrium~\cite{liu1970,reznik2011}.} \textcolor{black}{The value of $k_{\mathrm{bulk}}$ was not fitted to the experiments. It is of the same order as the seeded-growth rate measured by Liu and Nancollas at 25\,$^{\circ}$C with about $1\ \mathrm{g\,L^{-1}}$ of gypsum seed~\cite{liu1970}. Using a seeded-growth law means that nucleation and the induction time are not modelled: precipitation starts wherever $\mathrm{SR}>1$. Because the feeds contain no seed crystals, this is likely an upper estimate of bulk precipitation.} \textcolor{black}{With this value, bulk precipitation is slow compared with the passage of the fluid: for the reference case at $\mathrm{SR}=25$, the characteristic reaction time $C_{\mathrm{mix}}/(k_{\mathrm{bulk}}\sigma^{2})$ is about $10^{4}\ \mathrm{s}$, against about $8\ \mathrm{s}$ for the fluid to pass through the insert. The rate constants therefore scale the amount of gypsum formed, whereas the location of the deposit is set by where the two reactants meet the walls (\cref{sec:cfd}).} Over one time step, \cref{eq:bulkPrecipRate} was integrated explicitly as $\Delta C_{\mathrm{bulk}} = k_{\mathrm{bulk}}\sigma^{n_{\mathrm{bulk}}}\Delta t$ and applied stoichiometrically:

\begin{equation}
A^{n+1}=A^n-\Delta C_{\mathrm{bulk}},\qquad
B^{n+1}=B^n-\Delta C_{\mathrm{bulk}},\qquad
C^{n+1}=C^n+\Delta C_{\mathrm{bulk}},
\end{equation}

with the increment limited by the available reactants, $\Delta C_{\mathrm{bulk}} \le \min(A^n,B^n)$, to enforce non-negativity and local conservation.

A wall-associated deposition model was applied to the \texttt{flowCell} and \texttt{lid} patches, driven by $\mathrm{SR}$ in the wall-adjacent (owner) cell. The intrinsic wall flux was:

\begin{equation}
j_{\mathrm{kin}} = k_{\mathrm{wall}}\,\sigma_w^{\,n_{\mathrm{wall}}},
\qquad
\sigma_w = \max\!\left(\mathrm{SR}^{1/2} - X_{\mathrm{crit}}^{1/2},\,0\right),
\end{equation}

with $k_{\mathrm{wall}} = 1.0\times10^{-6}\ \mathrm{mol\,m^{-2}\,s^{-1}}$ and $n_{\mathrm{wall}}=2.3$. \textcolor{black}{Growth coefficients measured on gypsum surfaces at 25\,$^{\circ}$C are about $3\times10^{-5}$ to $1\times10^{-4}\ \mathrm{mol\,m^{-2}\,s^{-1}}$ at the salinity of the mixed feeds~\cite{he1994,reznik2011}. The lower value of $k_{\mathrm{wall}}$ represents a wall that is initially bare and only partly covered by growing crystals; it was likewise not fitted. The exponent lies between the second order found close to equilibrium and the third to fifth order found at higher supersaturation~\cite{witkamp1990}. Both rate laws are applied at $\sigma$ up to about 4, well beyond the range $\sigma\lesssim1$ in which they were measured.} To prevent unphysical wall uptake, the kinetic flux was limited by a diffusive supply estimate from the owner cell:

\begin{equation}
j = \min\!\left(j_{\mathrm{kin}},\,D_A\rho_wA\,\delta_f,\,D_B\rho_wB\,\delta_f\right),
\label{eq:wallsink}
\end{equation}

where $\rho_w=1000\ \mathrm{kg\,m^{-3}}$ and $\delta_f$ is the reciprocal wall-normal cell distance (\texttt{deltaCoeffs}) for face $f$. For a wall face of area $A_f$ adjacent to a cell of volume $V_c$, the deposited moles $\Delta n_f = jA_f\Delta t$ were converted to a decrease in molality $\Delta C_{\mathrm{wall}} = jA_f\Delta t/(\rho_wV_c)$ and applied stoichiometrically to the owner cell, again limited by $\min(A^n,B^n)$. Surface deposition is thus represented as a wall-adjacent cell sink rather than an explicit flux boundary condition.

The cumulative deposition per wall face was tracked as $N_{\mathrm{dep}}(t)=\int_0^t j\,\mathrm{d}t$, with $j$ from \cref{eq:wallsink}, and converted to an equivalent thickness of gypsum $h_{\mathrm{dep}}=N_{\mathrm{dep}}M_{\mathrm{gyp}}/\rho_{\mathrm{gyp}}$ using \textcolor{black}{the molar mass and density of gypsum, the dihydrate CaSO$_4\cdot2$H$_2$O,} $M_{\mathrm{gyp}}=0.17217\ \mathrm{kg\,mol^{-1}}$ and $\rho_{\mathrm{gyp}}=2320\ \mathrm{kg\,m^{-3}}$. These were diagnostic only: the deposit did not modify the mesh, wall position, or momentum equations.

\subsection{Geometry and mesh generation}
\label{sec:mesh}
A background hexahedral mesh was created with \texttt{blockMesh}, spanning $x\in[-2,\,440.814]$, $y\in[-2,\,25.25]$ and $z\in[-2,\,17]\ \mathrm{mm}$, discretized by a single structured block of $220\times20\times14$ cells with uniform grading. Body-fitted meshing used \texttt{snappyHexMesh} with four triangulated surfaces: \texttt{Cell\_flowchannel\_v2.stl} (patch \texttt{flowCell}, wall), \texttt{Lid.stl} (\texttt{lid}, wall), \texttt{Inlets.stl} (\texttt{inlets}, patch) and \texttt{Outlet.stl} (\texttt{outlet}, patch).

The inlet surface comprises two circular openings at $x=0$ with centers at $(y,z)=(8.06,\,11.6)$ and $(15.19,\,11.6)\ \mathrm{mm}$ and radius $r=1.6\ \mathrm{mm}$; the outlet is a single circular opening at $x=438.8135\ \mathrm{mm}$ with center $(y,z)=(11.625,\,8.0)\ \mathrm{mm}$ and radius $r=5.0\ \mathrm{mm}$. The two inlet openings were separated into patches \texttt{inletA} and \texttt{inletB} using \texttt{topoSet} and \texttt{createPatch} with box selections.

Meshing was executed with \texttt{castellatedMesh} and \texttt{snap} enabled and layer addition disabled. Surface refinement was applied at levels $(1,1)$ for \texttt{flowCell} and \texttt{lid} and $(2,2)$ for \texttt{inlets} and \texttt{outlet}, with \texttt{resolveFeatureAngle 30}. Mesh-quality controls were \texttt{maxNonOrtho 65} (relaxed to 75), \texttt{maxBoundarySkewness 4} and \texttt{maxInternalSkewness 4}.

The mesh dependence was assessed for the reference case with DDES-SA on five meshes of $6.1\times10^4$ to $6.7\times10^5$ cells, refined locally around the confluence, the step, the shear layer, the reattachment region and the side walls (Supplementary Material, Section~S2). The pressure drop and the \textcolor{black}{reattachment length differ} by less than 0.5 and $1.5\,\%$, respectively, between the two finest meshes ($3.4\times10^5$ and $6.7\times10^5$ cells), while the production mesh underestimates the pressure drop by about $40\,\%$ and the cross-channel velocity near the step by an order of magnitude. However, the reattachment length of the production RANS case, $9.2\ \mathrm{mm}$, lies within the range of the refined meshes ($9.2$ to $11.5\ \mathrm{mm}$). On the refined meshes, a small fraction of the Ca$^{2+}$ stream reaches the opposite side wall (on average $4\,\%$ of the inlet concentration in the near-wall region, compared with none on the production mesh). The reacting calculation was not repeated on a refined mesh.

Three geometries were used for the reacting campaign, with expansion ratios $E=H/h=1.42$, $2.06$ and $2.5$ (steps of $2.5$, $6.37$ and $9.0\ \mathrm{mm}$, measured on the meshes). The first corresponds to the \textcolor{black}{S2.5} insert; no S0 geometry was simulated. Each geometry was run at seven flow levels, with velocities of $0.0175$ to $0.4634\ \mathrm{m\,s^{-1}}$ at the circular inlet openings, and eight target saturation ratios, $\mathrm{SR}=0.47$ to $131$, giving in total 168 cases. The reaction was switched on at $t=10\ \mathrm{s}$ using the converged frozen flow field. All cases were run to $t=20\ \mathrm{s}$, 112 continued to $50\ \mathrm{s}$ and $84$ to $65\ \mathrm{s}$. \textcolor{black}{Cases with $\mathrm{SR}>42$ would require a Na$_2$SO$_4$ feed above its solubility; they were computed but are excluded from the analysis.} The reference case is the \textcolor{black}{S2.5} geometry at $48.5\ \mathrm{mL\,min^{-1}}$ per inlet and $\mathrm{SR}=25$.

\subsection{Initial and boundary conditions}
\label{sec:bcs}
The flow started from rest. Uniform velocities were imposed at both inlets, with zero-gradient velocity and fixed pressure at the outlet and no-slip walls. Turbulence fields were initialized at $k=0.01\ \mathrm{m^2\,s^{-2}}$ and $\varepsilon=0.1\ \mathrm{m^2\,s^{-3}}$, fixed at the inlets and treated with standard wall functions (\texttt{kqRWallFunction}, \texttt{epsilonWallFunction}, \texttt{nutkWallFunction}).

The scalar fields were initially set to zero ($A=B=C=0$). Inlet A supplied only Ca$^{2+}$ and inlet B only SO$_4^{2-}$, with inlet concentrations $A=C_{\mathrm{mix}}/f_A$ and $B=C_{\mathrm{mix}}/f_B$, respectively. The parameters $f_A$ and $f_B$ are the inlet area fractions, while $C_{\mathrm{mix}}$ denotes the equimolar concentration required to achieve the target saturation ratio on the \textcolor{black}{equimolar diagonal of the table, where the Ca$^{2+}$ and SO$_4^{2-}$ molalities are equal}. For the reference case, $\mathrm{SR}=25$, this gives $A=B=0.6065\ \mathrm{mol\,kg_w^{-1}}$, close to the feed molalities of the $0.3\ \mathrm{M}$ series in \cref{tab:density} ($0.593$ and $0.588\ \mathrm{mol\,kg_w^{-1}}$). All scalars had zero-gradient conditions at the outlet and walls.

\subsection{Discretization and solution control}
\label{sec:numerics}
The equations were discretized with the finite-volume method, using implicit Euler time integration and Gauss linear gradients. Convection used \texttt{linearUpwind} for momentum and first-order upwind for $k$, $\varepsilon$ and the scalars; Laplacian terms used \texttt{Gauss linear corrected} \cite{greenshields2025,fvSchemesGuide}. \textcolor{black}{First-order upwinding of the scalars adds numerical diffusion wherever the flow crosses the mesh obliquely, near the inlets, the confluence and the step, and broadens the simulated interface beyond what molecular diffusion alone would give. The simulated lateral transport of the reactants is therefore, if anything, overestimated, so the absence of Ca$^{2+}$ at the opposite side wall in the constant-density simulations (\cref{sec:cfd}) is a conservative result.} Pressure-velocity coupling used the PIMPLE algorithm with two correctors. Pressure was solved with GAMG and a \texttt{DICGaussSeidel} smoother (tolerance $10^{-7}$, relative tolerance $0.01$, and $0$ on the final corrector), $\mathbf{U}$, $k$ and $\varepsilon$ with a symmetric Gauss-Seidel smoother (tolerance $10^{-5}$, relative tolerance $0.1$), and the scalars to a tolerance of $10^{-8}$.

\section{Dimensionless groups}
\label{sec:dimensionless}
The simulations are characterized by the Reynolds number of the inlet channel, \textcolor{black}{$\mathrm{Re}_{D}=U_{\mathrm{in}}D/\nu$}, with the bulk velocity $U_{\mathrm{in}}$ \textcolor{black}{of the upstream channel and $D=2h$, the convention used for backward-facing steps~\cite{Pereira1983,Biswas2004} (the hydraulic diameter of the upstream duct, $2Wh/(W+h)$, is $0.68\,D$)}, and by the expansion ratio $E=H/h=1+S/h$. For experiments, we use height-based $\mathrm{Re}_H=UH/\nu$ downstream of the step; with $U=2Q/(WH)$ this is $\mathrm{Re}_H=2Q/(W\nu)$ and does not depend on $H$. Reynolds numbers use the kinematic viscosity of water, $\nu=1.004\times10^{-6}\ \mathrm{m^2\,s^{-1}}$ at $20\ ^{\circ}\mathrm{C}$ for the experiments and $8.9\times10^{-7}\ \mathrm{m^2\,s^{-1}}$ at $25\ ^{\circ}\mathrm{C}$ for the simulations~\cite{kestin1978}. The salt solutions are about 10\,\% more viscous than water (e.g.\ Na$_2$SO$_4$~\cite{abdulagatov2005}), so these Reynolds numbers are upper estimates by a few per cent.

The density \textcolor{black}{difference} between the streams is expressed through the reduced gravity, the buoyancy velocity, and the bulk Richardson number\textcolor{black}{~\cite{turner1973}}:

\begin{equation}
g' = g\,\frac{\Delta\rho}{\rho_0},
\qquad
U_b = \sqrt{g'H},
\qquad
Ri = \frac{g'H}{U^2} = \left(\frac{U_b}{U}\right)^{\!2},
\label{eq:richardson}
\end{equation}

with $\rho_0=998.2\ \mathrm{kg\,m^{-3}}$. $Ri$ compares the buoyancy force with the inertia of the imposed flow. The usual reason for neglecting buoyancy is that $\Delta\rho/\rho_0\ll1$, but $Ri$ also contains the channel height and the inverse square of the velocity. At low flow rates in the flow cell, a density difference that is small compared to $\rho_0$ can therefore still produce a buoyancy velocity larger than the imposed flow.
\section{Results}
\subsection{A constant-density simulation places deposition on the floor}
\label{sec:cfd}

\begin{figure*}[htbp]
    \centering
    \includegraphics[width=\linewidth]{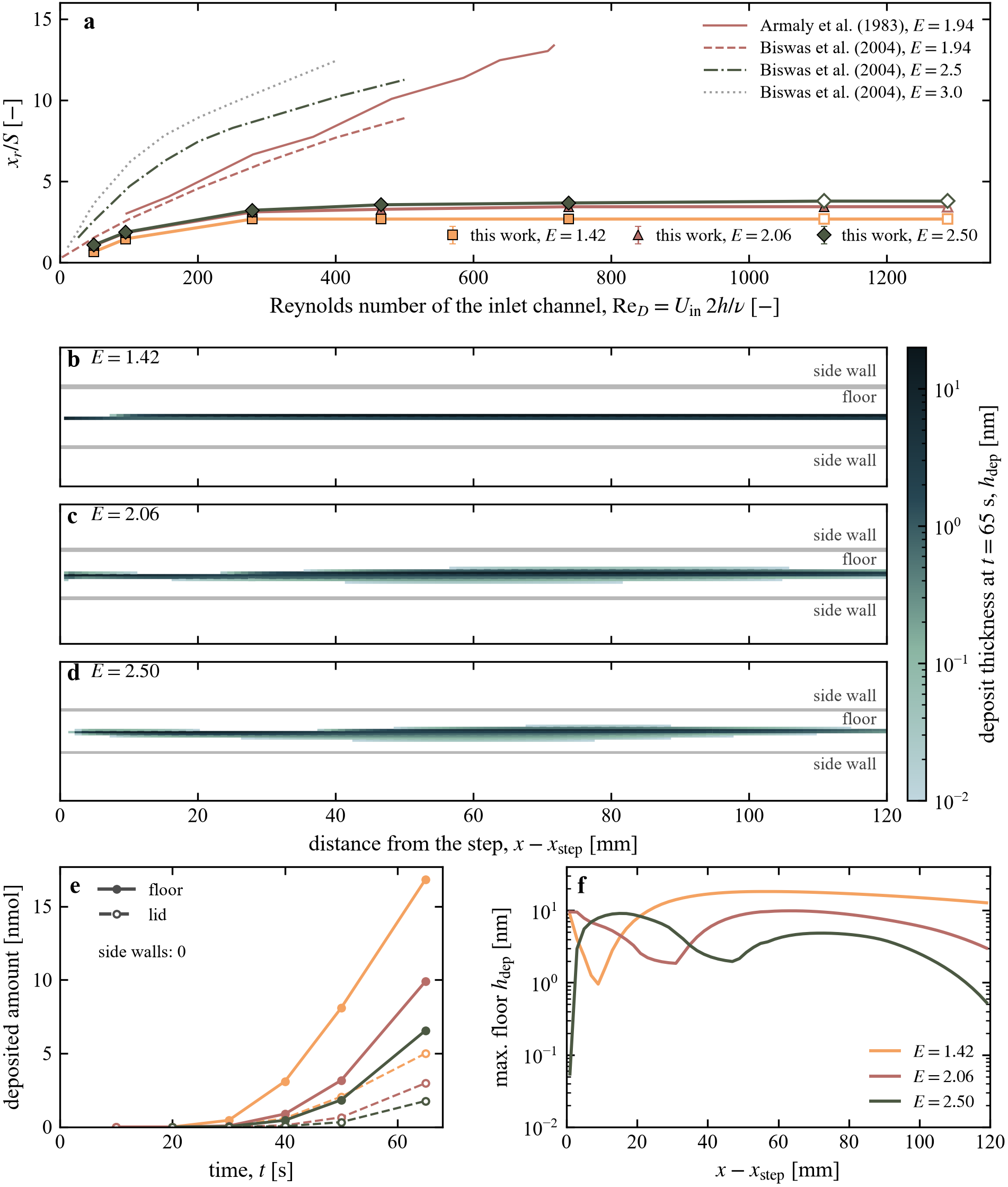}
    \caption{Constant-density simulations.
        \textbf{a} Reattachment length, normalized by the step height, against the Reynolds number of the inlet channel based on $D=2h$, for three expansion ratios, with published data \cite{Pereira1983,Biswas2004}. Error bars mark the resolution of the wall grid ($\pm0.5\ \mathrm{mm}$); open symbols are runs that had not reached a steady state. Colour marks the expansion ratio in panels a, e and f, including the published curves; marker shapes follow the channel depth used in \cref{fig:velocity,fig:collapse} (square $H=8.5\ \mathrm{mm}$, triangle $12.3\ \mathrm{mm}$), with a diamond for $H=15\ \mathrm{mm}$.
        \textbf{b}--\textbf{d} Gypsum deposit accumulated by $t=65\ \mathrm{s}$ on the floor and the two side walls (folded out about the floor edges) over the $120\ \mathrm{mm}$ insert, for $E=1.42$ (\textcolor{black}{S2.5}), 2.06 and 2.5 at $48.5\ \mathrm{mL\,min^{-1}}$ per inlet and $\mathrm{SR}=25$. \textcolor{black}{White areas indicate deposits thinner than} $0.01\ \mathrm{nm}$.
        \textbf{e} Deposited amount within the insert on the floor and the lid against time.
        \textbf{f} Largest deposit thickness across the floor, in $2\ \mathrm{mm}$ bins along the insert, at $t=65\ \mathrm{s}$.}
    \label{fig:cfd}
\end{figure*}

We first compare the constant-density simulations with the published results for flow over a backward-facing step (\cref{fig:cfd}a). Note that we have normalized the studies using $x_r/S$ against $\mathrm{Re}_D=\textcolor{black}{U_{\mathrm{in}}}\,2h/\nu$ (\cref{sec:dimensionless}). For all three expansion ratios, the reattachment length increases with Reynolds number and then levels off above $\mathrm{Re}_D\approx300$ to 500, at $x_r/S\approx2.7$ for $E=1.42$, $3.4$ for $E=2.06$ and $3.8$ for $E=2.5$. While the ordering with expansion ratio is similar to the published data, the obtained \textcolor{black}{reattachment lengths} are significantly lower than the published data. Both Armaly et al.\ \cite{Pereira1983} and Biswas et al. \cite{Biswas2004} report a reattachment length that keeps rising, to 9 to 13 step heights at $\mathrm{Re}_D=500$ to 700. Two aspects of the calculation limit this comparison. First, the reattachment point is identified on the wall-adjacent grid cells, giving a resolution of approximately $1\ \mathrm{mm}$, which corresponds to the size of the error bars. Second, the computational domain is relatively narrow, with a width of only 2.2 upstream channel heights, whereas the reference data were obtained in two-dimensional or sufficiently wide channels, where side-wall confinement is negligible \cite{Pereira1983,Biswas2004}. Consequently, the comparison should be regarded as a qualitative assessment of the separated-flow behaviour rather than a validation of the model.

The reactive species were then transported on this constant-density flow, with the flow field held fixed and the reaction switched on at $t=10\ \mathrm{s}$. \Cref{fig:cfd}b-d show the gypsum that has accumulated on the walls by $t=65\ \mathrm{s}$ at the reference flow rate ($48.5\ \mathrm{mL\,min^{-1}}$ per inlet) and $\mathrm{SR}=25$. In all three geometries the deposit forms on the floor, along the channel centreline where the two feeds meet. In the \textcolor{black}{S2.5} geometry ($E=1.42$), the deposit forms a narrow strip only two wall-cell rows wide, located at $y=11.3$ and $12.0\ \mathrm{mm}$ on either side of the centreline at $11.6\ \mathrm{mm}$. Its width is therefore determined by the mesh resolution rather than the flow. The deposit reaches a thickness  of $18\ \mathrm{nm}$ at $65\ \mathrm{s}$ and extends along most of the insert (\cref{fig:cfd}f). When increasing the step height, the deposited strip broadens downstream of the recirculation zone to about $4\ \mathrm{mm}$ (based on faces exceeding $0.01\ \mathrm{nm}$ in thickness), while remaining  centred on the channel centreline. The resulting deposits are thinner, reaching thicknesses of only 9 to $10\ \mathrm{nm}$. Within the insert, 77 to $79\,\%$ of the deposited amount lies on the floor and 21 to $23\,\%$ on the lid, also along the centreline. There is no deposition on either side wall in any of the three geometries (\cref{fig:cfd}e). The deposition begins between 20 and $30\ \mathrm{s}$ and then accelerates; between 50 and $65\ \mathrm{s}$ the thickest point in the \textcolor{black}{S2.5} geometry grows at $0.54\ \mathrm{nm\,s^{-1}}$. Reducing the saturation ratio to 9.96 decreases the deposited amount at $65\ \mathrm{s}$ by an order of magnitude, whereas increasing it to 42 enhances deposition by a factor of 3.4. In all cases, the deposit \textcolor{black}{remains} confined to the centreline of the floor and lid of the channel.

The simulation thus predicts that gypsum should grow where the two reactant streams meet the floor and the lid, that is, along the line of contact between the two streams in the middle of the channel.

\subsection{In the experiment gypsum deposits on the side walls}
\label{sec:bands}

\begin{figure*}[htbp]
    \centering
    \includegraphics[width=\linewidth]{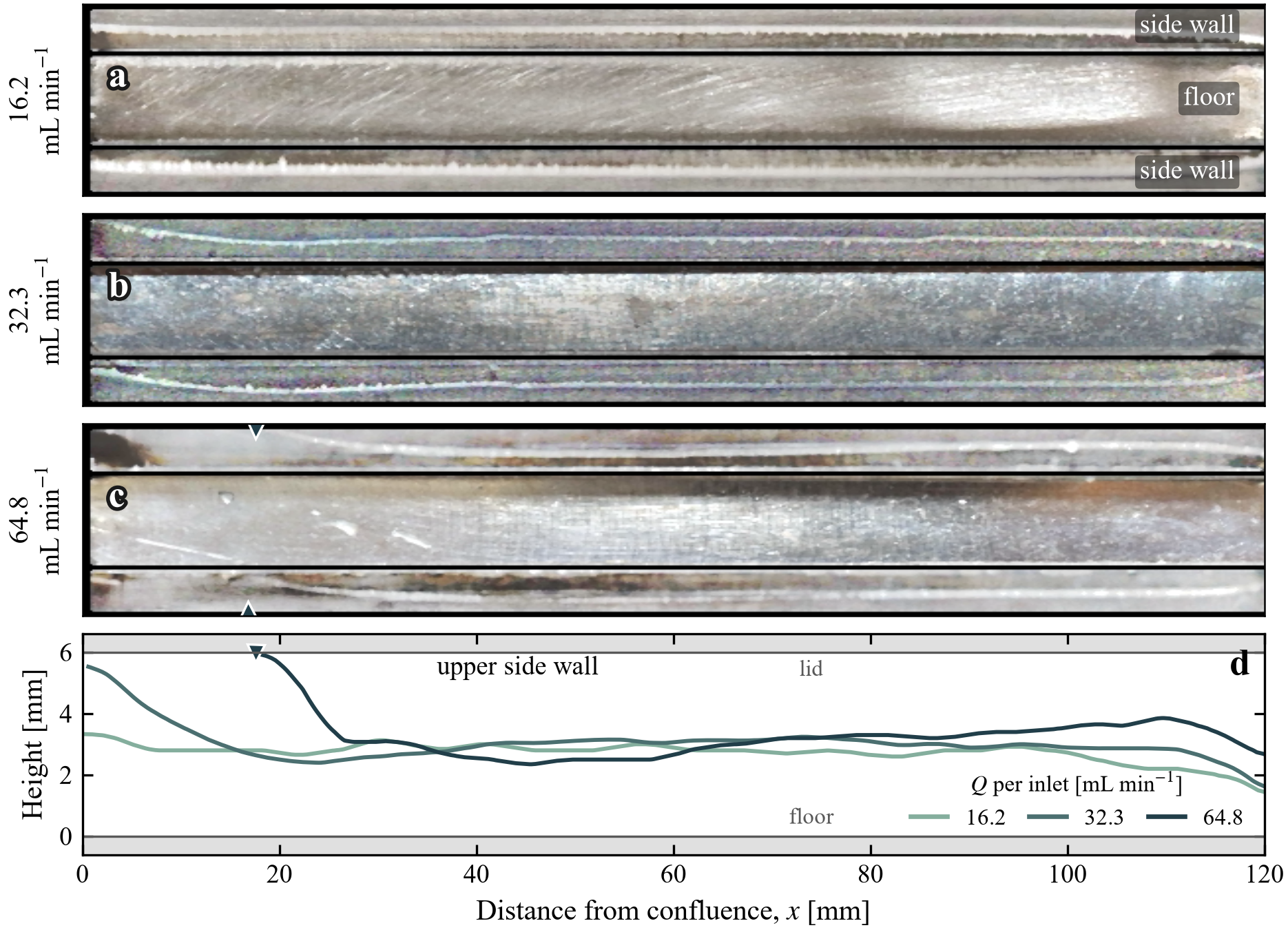}
    \caption{Gypsum deposition in the S0 configuration without density matching.
        \textbf{a}--\textbf{c} Unfolded views of the two side walls (top and bottom strips) and the floor coupon (middle strip) after runs at 16.2, 32.3 and $64.8\ \mathrm{mL\,min^{-1}}$ per inlet; flow from left to right, $x=0$ at the confluence. The white line on each side wall is the gypsum band; the floor carries no visible deposit. Colour noise was reduced with a median filter \textcolor{black}{applied to the colour channels only, leaving the brightness unchanged}.
        \textbf{d} Band centreline as height above the floor on the upper side wall of the unfolded view. Triangles mark where the band begins at $64.8\ \mathrm{mL\,min^{-1}}$; at the two lower flow rates it is present from the confluence. The grey strips lie outside the $6\ \mathrm{mm}$ channel.}
    \label{fig:band}
\end{figure*}

The reacting experiments did not reproduce the deposition patterns of the simulation. Instead, gypsum formed bands on both side walls, while no visible deposit was observed on the floor (\cref{fig:band}a--c). The most pronounced effect of the flow rate was on the location at which the bands start. At $64.8\ \mathrm{mL\,min^{-1}}$ per inlet, the bands began $17.6$ and $16.8\ \mathrm{mm}$ downstream of the confluence on the two side walls. Reducing the flow rate to $32.3\ \mathrm{mL\,min^{-1}}$ shifted the onset of deposition to within $0.6\ \mathrm{mm}$ of the confluence, while at $16.2\ \mathrm{mL\,min^{-1}}$ deposition was present from the confluence itself.

The bands are generally observed near the mid-height of the $6\ \mathrm{mm}$ channel. At the two higher flow rates, deposition \textcolor{black}{started} near the lid on one wall and near the floor on the other,  before converging to mid-height within 10 to $20\ \mathrm{mm}$ of the onset location. At $16.2\ \mathrm{mL\,min^{-1}}$, in contrast, the band was located near mid-height from the confluence (\cref{fig:band}d for the upper wall). Although the mean band height decreased slightly with flow rate, from $3.2$ to $3.0$ and $2.7\ \mathrm{mm}$ on one wall and from $3.6$ to $3.5$ and $3.0\ \mathrm{mm}$ on the other, this trend was observed at only about half of the streamwise positions along the channel. In addition, the run-to-run variation at $32.3\ \mathrm{mL\,min^{-1}}$ was approximately $0.2\ \mathrm{mm}$. We therefore consider the onset location, rather than the band height, to be the more robust experimental observation.

More gypsum formed on the wall adjacent to the Na$_2$SO$_4$ inlet than on the CaCl$_2$ side, and the asymmetry persisted when the two pumps were exchanged. The pumps and tubing can therefore be excluded as the source of the effect. Density differences, however, remain a plausible explanation, since exchanging the pumps does not alter the side of the cell on which each \textcolor{black}{solution} enters. Gypsum also grew on adhesive tape applied to one side wall, indicating that the wall material does not control where the deposit occurs.

The side-wall bands contradict the floor-centreline strip predicted by the simulation. \textcolor{black}{The simulations used the S2.5 geometry, whereas the reacting experiments used S0 (\cref{sec:mesh}); the difference in step height does not account for the discrepancy, because the predicted deposit stays on the floor and lid centreline in all three simulated geometries (\cref{sec:cfd}).} Neither \textcolor{black}{varying} the saturation ratio nor changing the flow rate displaced the \textcolor{black}{predicted deposit} from the centreline, and a different turbulence closure likewise failed to transport the Ca$^{2+}$ stream to the far wall at the reference flow rate. Scale-resolving simulations at constant density mix the streams faster once the flow becomes unsteady, above approximately $60\ \mathrm{mL\,min^{-1}}$ per inlet, but they do not reorient the interface across the channel height (\cref{sec:governing}; Supplementary Material, Section~S1). The discrepancy therefore appears to originate from the predicted transport of the reactants rather than from the deposition model itself. The following experiments examine this transport directly.

\subsection{The density difference controls how far the dyed stream spreads}
\label{sec:control}

\begin{figure*}[htbp]
    \centering
    \includegraphics[width=\linewidth]{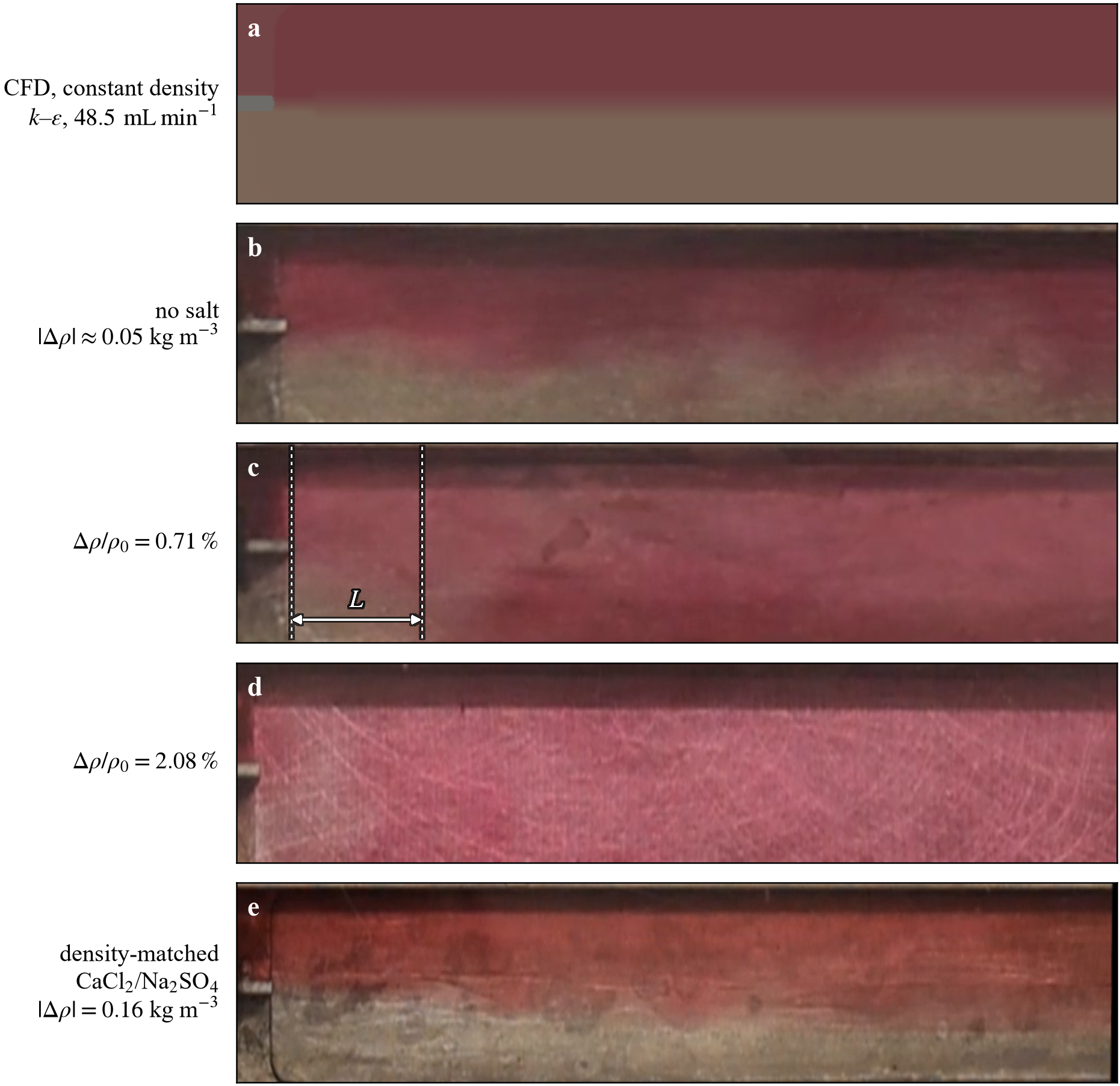}
    \caption{Top views at $50\ \mathrm{mL\,min^{-1}}$ per inlet; flow from left to right, dyed stream entering at the top.
        \textbf{a} Constant-density simulation ($k$--$\varepsilon$, reference case, $48.5\ \mathrm{mL\,min^{-1}}$ per inlet) rendered as the camera sees it: the passive tracer is integrated over the channel height and shaded between the clear-fluid and dye colours measured in panel b, with the optical depth taken relative to $8.5\ \mathrm{mm}$ of undiluted dye. Shown from 3 to $67\ \mathrm{mm}$ downstream of the splitter tip.
        \textbf{b}--\textbf{d} Water with NaCl added to the undyed stream (\textcolor{black}{S2.5} insert): no salt, $\Delta\rho/\rho_0=0.71\,\%$ and $2.08\,\%$. The dyed stream is the \textcolor{black}{less dense one}.
        \textbf{e} Density-matched CaCl$_2$/Na$_2$SO$_4$ feeds at a mixed concentration of $0.3\ \mathrm{M}$, with a residual difference of $0.16\ \mathrm{kg\,m^{-3}}$; here the dye is in the Na$_2$SO$_4$ stream. \textcolor{black}{In \textbf{c}, the arrow between the dotted lines marks the clear-zone length $L$: the distance from the confluence to where dye first reaches the lower wall, opposite the dyed inlet.}}
    \label{fig:density}
\end{figure*}

\Cref{fig:density} compares four dye experiments conducted at a flow rate of $50\ \mathrm{mL\,min^{-1}}$ per inlet with the predictions of the constant-density simulation reference case (\cref{fig:density}a). In the simulation, the dyed stream remains confined to one half of the channel, separated from the clear stream by a sharp, nearly straight interface along the centreline. The experiment without added salt resembles this behaviour: the two streams remain largely side by side throughout the channel, although the interface between them is no longer straight, because it is already unsteady at this flow rate. As the density difference increases, the dyed stream spreads progressively further \textcolor{black}{across} the channel. At $\Delta\rho/\rho_0=0.71\,\%$ the spreading is clearly enhanced, while at $2.08\,\%$ the dyed fluid reaches the full channel width within a short distance of the confluence. Because the camera integrates light over the full channel depth, these images do not necessarily indicate complete mixing. The images do demonstrate, however, that the \textcolor{black}{dyed} fluid reaches the opposite wall, in contrast to constant-density simulations.

One possible explanation for the dyed stream reaching the opposite wall is buoyancy induced flow arising from the density difference between two feeds. The magnitude of this effect can be estimated from the measured density difference. For the density difference $2.08\,\%$ in the $8.5\ \mathrm{mm}$ channel, the reduced gravity is $g'=0.204\ \mathrm{m\,s^{-2}}$, corresponding to a \textcolor{black}{buoyancy} velocity scale of $U_b=\sqrt{g'H}=41.6\ \mathrm{mm\,s^{-1}}$. This is  2.8 times larger than the imposed bulk velocity of $15.1\ \mathrm{mm\,s^{-1}}$. The density difference is therefore not a small \textcolor{black}{perturbation} to the flow: it corresponds to a bulk Richardson number of approximately $Ri=7.6$, indicating that \textcolor{black}{buoyancy forces are expected} to strongly influence transport. 

The density-matched situation provides a direct test of the role of \textcolor{black}{buoyancy} (\cref{tab:control}). In experiments using the reacting salt solutions, the dyed stream reached the opposite wall of the channel at every flow rate when the \textcolor{black}{density difference} was present, with median  lengths of the clear zone of  $18.3$ and $15.9\ \mathrm{mm}$ at mixed concentrations of $0.1$ and $0.3\ \mathrm{M}$. In contrast, when the densities were matched, the dyed stream did not reach the opposite wall within the $120\ \mathrm{mm}$ channel in \textcolor{black}{$78\,\%$ ($0.1\ \mathrm{M}$) and $62\,\%$ ($0.3\ \mathrm{M}$)} of the tested conditions\textcolor{black}{, i.e.\ in 25 and 20 of the 32 flow rates}. The cases that did reach the opposite wall fall into two categories. At the lowest flow rates, 5 to $15\ \mathrm{mL\,min^{-1}}$, the small residual density difference between the matched feeds (\cref{sec:fluids}) still gives $Ri\approx0.3$ to 2 in a $6\ \mathrm{mm}$ channel. The observed \textcolor{black}{clear-zone lengths, $L=9$ to $75\ \mathrm{mm}$ (the distance from the confluence to where dye is first detected at the opposite wall), are of the same order as the $13$ to $44\ \mathrm{mm}$ predicted for $Ri=2$ to $0.3$ by the correlation of \cref{sec:collapse} (\cref{eq:collapse})}. At flow rates of $65\ \mathrm{mL\,min^{-1}}$ and above, however, the interface becomes unsteady. In these experiments, unsteadiness \textcolor{black}{consistently} emerged at $\mathrm{Re}_H\approx107$, corresponding to approximately $42\ \mathrm{mL\,min^{-1}}$, irrespective of the density difference (Supplementary Material, Section~S5). \textcolor{black}{This onset lies below the 60 to $90\ \mathrm{mL\,min^{-1}}$ at which the constant-density LES becomes unsteady (\cref{sec:governing}). The LES has steady inlets, whereas each peristaltic pump delivers its flow in pulses at the pump frequency (about $4\ \mathrm{Hz}$ at $42\ \mathrm{mL\,min^{-1}}$), which perturb the streams at the confluence, so an earlier onset in the experiments is expected.} Under these flow conditions, flow-induced mixing transports the dye to the opposite wall even in the absence of significant \textcolor{black}{buoyancy} effects. Matching the feed densities required the addition of NaCl to one stream, so these controls do not completely exclude contributions from changes in viscosity, diffusivity or optical \textcolor{black}{response}. However, none of these effects would be expected to alter transport to the extent observed here, making \textcolor{black}{buoyancy} the most plausible explanation for the substantially different transport behaviour.

\begin{table}[htbp]
    \centering
    \caption{Clear-zone lengths in the reacting-salt dye experiments with and without density matching, over 32 flow rates from 5 to $160\ \mathrm{mL\,min^{-1}}$ per inlet. Conditions \textcolor{black}{in which the dye did not reach the opposite wall are assigned} $L=120\ \mathrm{mm}$, so the matched medians are lower bounds.}
    \label{tab:control}
    \begin{tabular}{lcccc}
        \toprule
        & \multicolumn{2}{c}{$0.1\ \mathrm{M}$} & \multicolumn{2}{c}{$0.3\ \mathrm{M}$} \\
        \cmidrule(lr){2-3}\cmidrule(lr){4-5}
        & unmatched & matched & unmatched & matched \\
        \midrule
        $|\Delta\rho|\ [\mathrm{kg\,m^{-3}}]$ & $7.37$ & $0.19$ & $20.04$ & $0.16$ \\
        Median clear-zone length $[\mathrm{mm}]$ & $18.3$ & $>120$ & $15.9$ & $>120$ \\
        Conditions without crossing & $0\,\%$ & $78\,\%$ & $0\,\%$ & $62\,\%$ \\
        \bottomrule
    \end{tabular}
\end{table}

\subsection{The front speed depends on the density difference, not on the flow rate}
\label{sec:velocity}

\begin{figure*}[htbp]
    \centering
    \includegraphics[width=\linewidth]{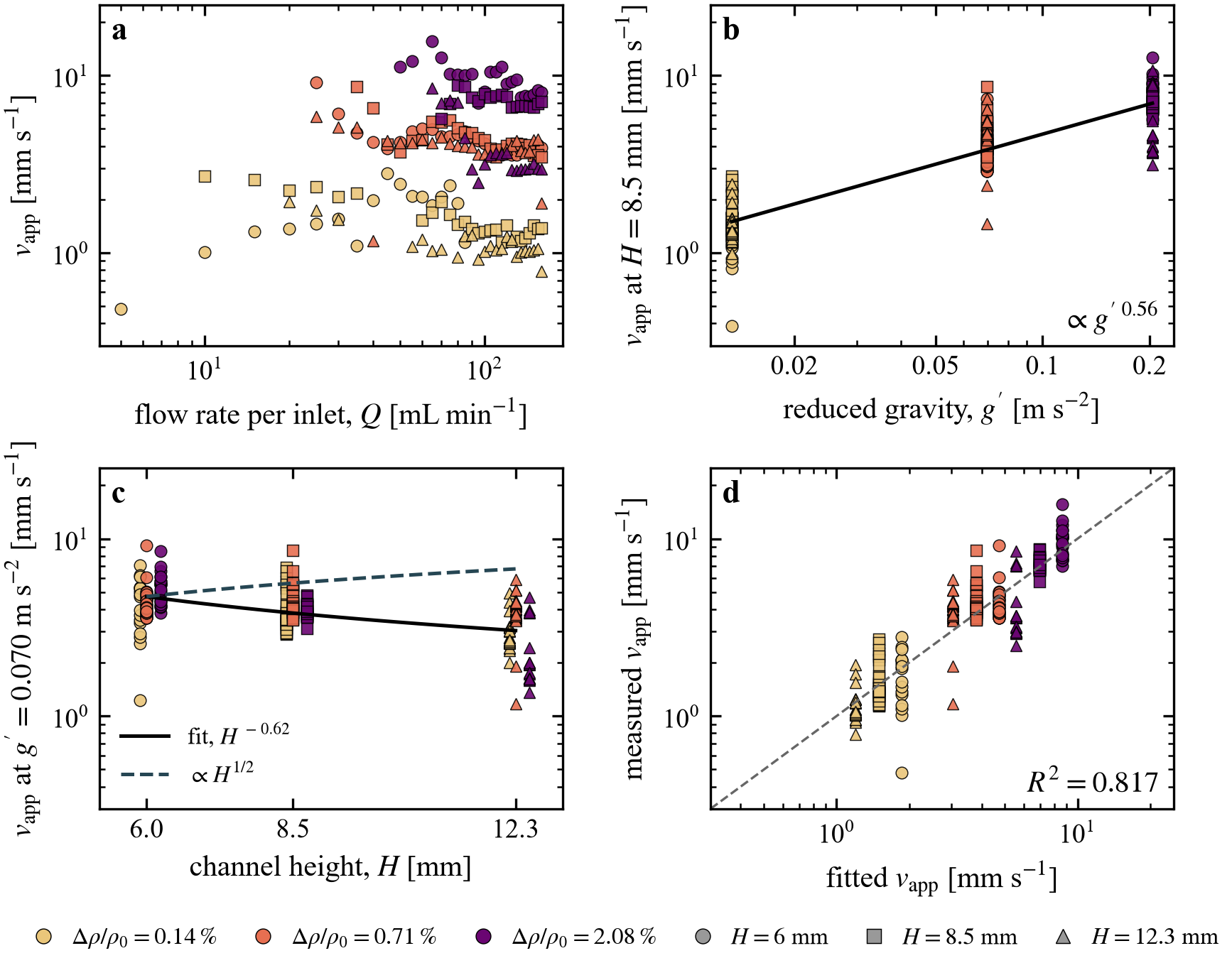}
    \caption{Apparent speed of the dye front towards the far wall.
        \textbf{a} Measured speed against flow rate per inlet.
        \textbf{b} Speed referred to $H=8.5\ \mathrm{mm}$ with the fitted height dependence, against reduced gravity.
        \textbf{c} Speed referred to $g'=0.070\ \mathrm{m\,s^{-2}}$ with the fitted $g'$ dependence, against channel height; the dashed line is the $H^{1/2}$ dependence of a gravity current with constant Froude number.
        \textbf{d} Measured against fitted speed, \cref{eq:velocity}.
        Colour gives the density difference and symbol shape the channel height.}
    \label{fig:velocity}
\end{figure*}

If buoyancy drives the dyed stream across the channel, the associated \textcolor{black}{transport velocity should depend} primarily on the density difference rather than on the imposed flow rate. The dye experiments with non-zero density \textcolor{black}{difference} span three channel heights, three density differences, and 32 flow rates, yielding 288 conditions in total. In 211 of these cases, the \textcolor{black}{interface between the clear and dyed fluid could be traced from the confluence to the opposite wall (200 cases) or to the outlet (11 cases)} \textcolor{black}{(the cross-sections in \cref{fig:fullcircle}b sketch such a progression at successive downstream positions)}. The resulting apparent velocity, $v_{\mathrm{app}}$, shows little dependence \textcolor{black}{on} flow rate at fixed density difference and height, but increases with increasing density difference (\cref{fig:velocity}a). A regression over the tracked conditions yields:

\begin{equation}
    v_{\mathrm{app}} \propto g'^{\,0.564}H^{-0.615},
    \qquad R^2 = 0.817 .
    \label{eq:velocity}
\end{equation}

\textcolor{black}{Introducing} the bulk velocity as an additional predictor \textcolor{black}{produces} an exponent of $-0.14$, with a bootstrap $95\,\%$ interval of $[-0.29,\,0.01]$, and increases $R^2$ by only 0.012. The apparent front velocity therefore depends only weakly, if at all, on the imposed flow rate, while scaling with the density difference through an exponent close to one half (\cref{fig:velocity}b). This is \textcolor{black}{consistent with transport driven by} buoyancy.

The dependence on channel height is the opposite of that expected for a gravity current propagating at constant Froude number. Under such a scaling $v_{\mathrm{app}}\propto\sqrt{g'H}$, \textcolor{black}{implying} that the propagation speed should increase with the \textcolor{black}{channel} height as $H^{1/2}$. Instead the measured speed decreases with increasing height (\cref{fig:velocity}c). Pairwise comparisons of conditions with the same  density difference and flow rate \textcolor{black}{but} different channel heights yield the same trend, with a median exponent of $-0.45$. Consistent with this observation, the \textcolor{black}{buoyancy} velocity scale $\sqrt{g'H}$ alone provides a relatively poor description of the data ($R^2=0.595$), whereas $\sqrt{g'/H}$ performs nearly as well as the unconstrained fit ($R^2=0.816$). Because $\sqrt{H/g'}$ corresponds to the characteristic time required for dense fluid to traverse the channel height under \textcolor{black}{buoyancy}, one \textcolor{black}{interpretation} is that this vertical time scale controls the apparent lateral advance of the dyed front. \textcolor{black}{Dividing the front speed by the buoyancy rate $\sqrt{g'/H}$ gives a characteristic length $v_{\mathrm{app}}/\sqrt{g'/H}$, whose mean values are $1.33$, $1.41$ and $1.17\ \mathrm{mm}$ for the three channel heights.} In addition, normalizing the data by $\sqrt{g'/H}$ reduces the coefficient of variation from 0.43 to 0.30, indicating a substantially improved collapse. The projected images, however, cannot distinguish vertical motion from lateral spreading. In addition, a front of given thickness is harder to detect in a deeper channel. Therefore, we do not assign a physical interpretation to the measured height exponent beyond noting that it is inconsistent with the classical gravity-current scaling.

\subsection{The Richardson number organizes the clear-zone length}
\label{sec:collapse}

\begin{figure*}[htbp]
    \centering
    \includegraphics[width=\linewidth]{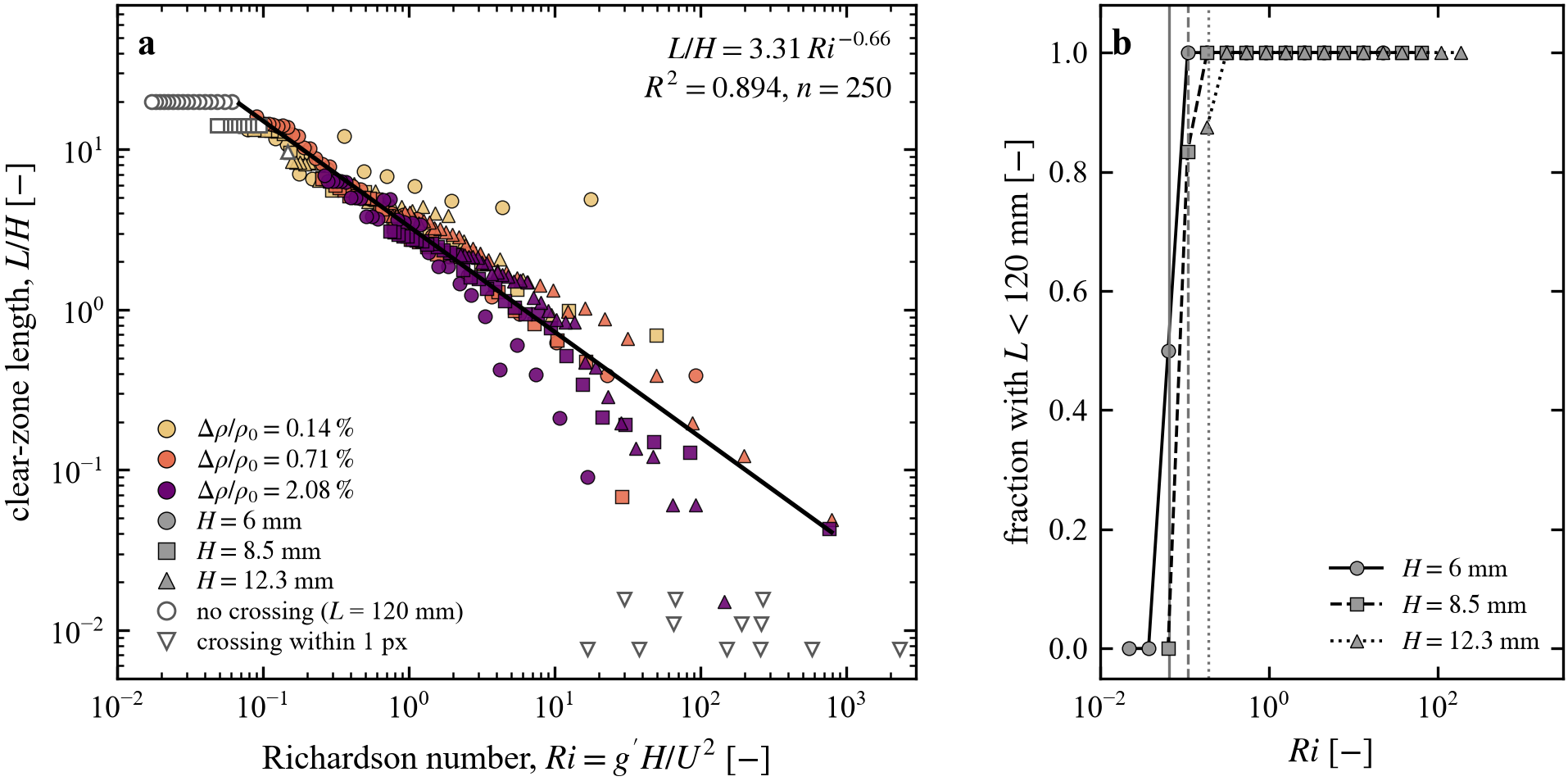}
    \caption{Clear-zone length and Richardson number.
        \textbf{a} Clear-zone length normalized by channel height against $Ri$, with the fit of \cref{eq:collapse}. Open circles and squares: no crossing within the channel ($L=120\ \mathrm{mm}$); open triangles: crossing within one pixel, drawn at half the pixel size.
        \textbf{b} Fraction of conditions with a crossing within the channel, per channel height. Vertical lines mark the $Ri$ at which \cref{eq:collapse} gives $L=120\ \mathrm{mm}$ for each height.}
    \label{fig:collapse}
\end{figure*}

The density difference alone does not decide how far downstream the dyed fluid reaches the opposite wall; the flow rate and the channel height are also important. These effects are combined in the bulk Richardson number. For the 250 conditions in which the dyed stream \textcolor{black}{crossed} the channel within the observation length, the length of the clear zone collapses onto
\begin{equation}
    \frac{L}{H}=3.31\,Ri^{-0.66},
    \qquad R^2=0.894,
    \label{eq:collapse}
\end{equation}
over four decades of $Ri$ from $0.068$ to 789 (\cref{fig:collapse}a). The bootstrap $95\,\%$ interval on the exponent is $[-0.70,\,-0.62]$. No other single parameter provides a comparable description of the data: \textcolor{black}{power-law fits} based on bulk velocity, the flow rate, and the reduced gravity \textcolor{black}{yield} $R^2$ values of only $0.50$, $0.39$ and $0.22$, respectively. The Richardson number combines the effects of a fifteen-fold variation in density difference and a twofold variation in channel height. To test whether the collapse is an artefact of pooling data from different channel heights, the analysis was repeated for each height separately. The resulting exponents, $-0.656$, $-0.668$ and $-0.688$ (Supplementary Material, Section~S4), all lie within the confidence interval of the pooled fit. The observed collapse therefore reflects \textcolor{black}{Richardson-number scaling rather} than an effect of combining data from different channel heights.

Two groups of conditions are excluded from \cref{eq:collapse}. In 26 cases, the dyed stream did not reach the opposite wall within the $120\ \mathrm{mm}$ observation length, \textcolor{black}{so only a lower bound on $L$ is known; these were recorded as $L=120\ \mathrm{mm}$ and excluded from the fit}. In another 12 cases, all at $Ri\geq17$, the dyed stream reached the opposite wall within one pixel \textcolor{black}{($0.18$ to $0.31\ \mathrm{mm}$)} of the confluence, so that the measured length of the clear zone was effectively zero; therefore, \textcolor{black}{only an upper bound on $L$ is known for these cases, and they were also excluded from the fit}. The fitted exponent is sensitive to the \textcolor{black}{treatment} of the latter group. \textcolor{black}{Assigning} these observations a length equal to the resolution limit \textcolor{black}{steepened} the scaling exponent from -0.66 to $-0.77$ ($R^2=0.850$), a change that exceeds the bootstrap 95\% confidence interval of the original fit. Consequently, the  exponent reported in \cref{eq:collapse} should be interpreted as describing the resolved range of data and, if anything, as a lower bound on the true steepness of the Richardson number dependence.

The fraction of conditions for which the dyed stream reaches the opposite wall within the channel decreases sharply at low $Ri$ (\cref{fig:collapse}b). This transition occurs where \cref{eq:collapse} predicts a length of the clear zone equal to the channel length. The corresponding Richardson numbers are \textcolor{black}{$0.065$}, $0.110$ and $0.194$ for the three heights, compared with the measured values of $0.068$, $0.094$ and $0.147$ at which half of the conditions \textcolor{black}{exhibit} crossing. The close agreement indicates that the decline in crossing \textcolor{black}{probability} is determined primarily by the finite  channel length rather than by a change in the underlying flow behaviour. Within the \textcolor{black}{experimentally} considered range, decreasing $Ri$ \textcolor{black}{therefore} does not eliminate the cross channel transport; it simply shifts the location at which the dyed fluid reaches the opposite \textcolor{black}{wall farther} downstream.

\subsection{With matched densities, gypsum deposits on the floor}
\label{sec:fullcircle}

\begin{figure*}[htbp]
    \centering
    \includegraphics[width=0.86\linewidth]{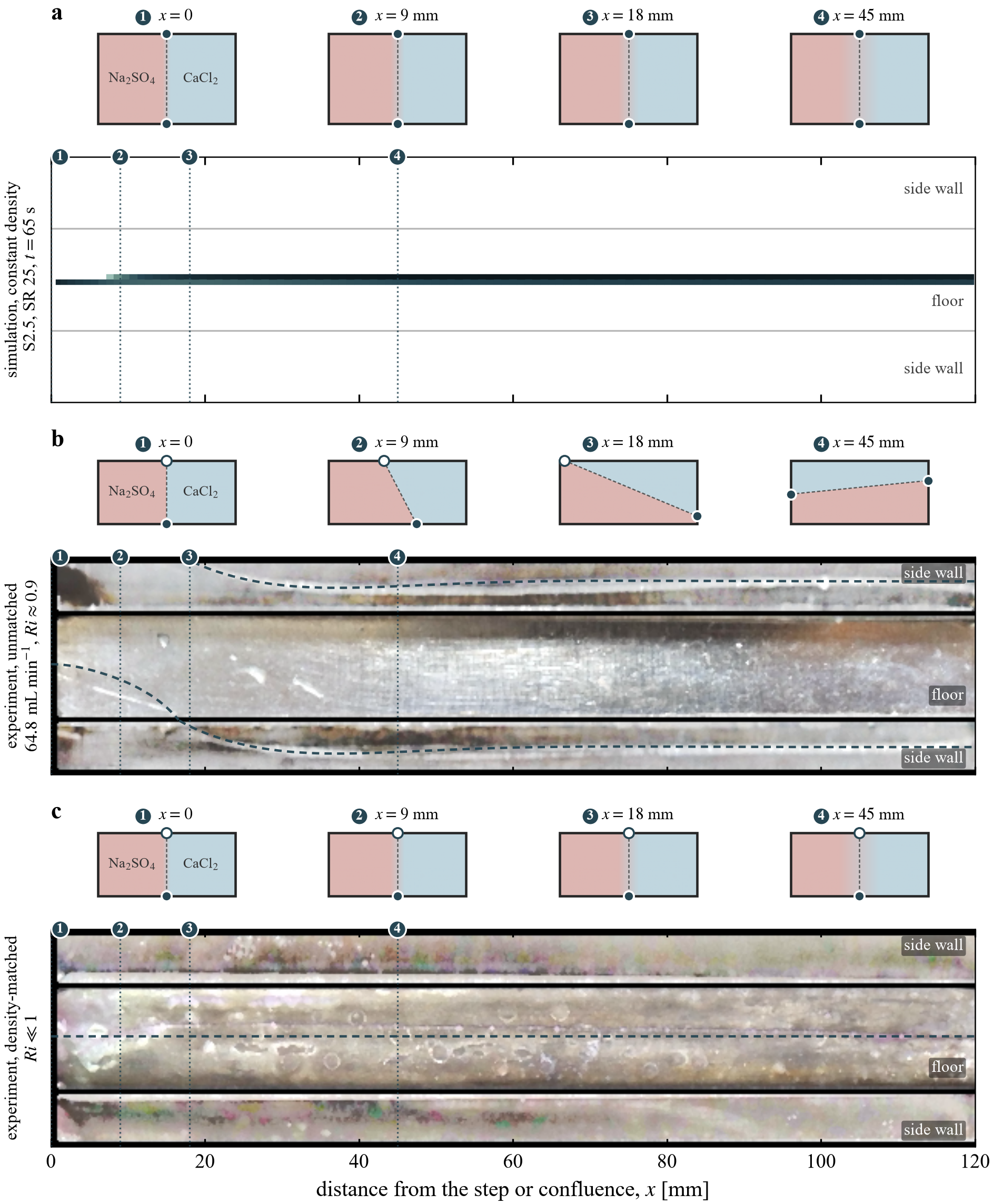}
    \caption{Where the interface between the two feeds meets the walls, and where gypsum is found. \textbf{a} Constant-density simulation (S2.5, $\mathrm{SR}=25$, $48.5\ \mathrm{mL\,min^{-1}}$ per inlet): deposit at $t=65\ \mathrm{s}$; colour scale as in \cref{fig:cfd}. \textbf{b} Reacting experiment, unmatched feeds (S0, $64.8\ \mathrm{mL\,min^{-1}}$ per inlet, $Ri\approx0.9$, after $25\ \mathrm{min}$), unfolded as in \cref{fig:band}. \textbf{c} Reacting experiment, densities matched with NaCl in the CaCl$_2$ feed (S0, $0.3\ \mathrm{M}$ mixed, $50\ \mathrm{mL\,min^{-1}}$ per inlet, $Ri\approx0.01$, after $15\ \mathrm{min}$). Cross-sections look downstream at the numbered positions, with the denser Na$_2$SO$_4$ feed on the side of the upper strip; dots mark where the interface meets a wall (open dots: lid, not imaged). The tilt in \textbf{b} is fitted to the band centrelines (\cref{sec:fullcircle}), not computed; in \textbf{a} and \textbf{c} the interface is drawn vertical.}
    \label{fig:fullcircle}
\end{figure*}

If buoyancy is responsible for transporting the reactants to the side walls, removing the density difference should \textcolor{black}{shift the} deposit back towards \textcolor{black}{the} location predicted by the constant-density simulation. Consistent with this expectation, a reacting experiment in which the feed densities were matched by adding NaCl to the \textcolor{black}{less dense} solution showed no side-wall bands; instead, gypsum deposited as a narrow line on the channel floor (\cref{fig:fullcircle}). For the density-matched experiments at $50\ \mathrm{mL\,min^{-1}}$ in the S0 insert, the residual difference of about $0.1\ \mathrm{kg\,m^{-3}}$ corresponds to $Ri\approx0.01$, compared with $Ri\approx2.6$ for unmatched feeds under otherwise identical conditions. In the unfolded view (\cref{fig:fullcircle}c), the deposited line first appears about $14\ \mathrm{mm}$ downstream of the confluence. It is \textcolor{black}{displaced} by up to $2\ \mathrm{mm}$ from the  floor centreline near the onset, but gradually returns towards the centreline and is centred by $x\approx110\ \mathrm{mm}$. No deposition bands are \textcolor{black}{observed} on either side wall. Thus\textcolor{black}{,} when the density difference is removed, deposition occurs on the same surface predicted by the simulations (\cref{fig:fullcircle}a) and in a similarly narrow configuration. Although the width, exact position and deposited amount have not been compared quantitatively, the qualitative agreement indicates that \textcolor{black}{buoyancy}-driven transport, rather than \textcolor{black}{the} deposition model itself, is responsible for the side wall deposition.

\Cref{fig:fullcircle} also illustrates the mechanism proposed in \cref{sec:disc_mechanism}. If the interface between the feeds rotates from vertical towards horizontal as the denser Na$_2$SO$_4$ feed underflows, the lines where it meets the walls move from the floor and lid onto the two side walls: onto one side wall from the lid and onto the other from the floor. A single tilting interface describes both measured band centrelines at $64.8\ \mathrm{mL\,min^{-1}}$ (\cref{fig:fullcircle}b) with a root mean square deviation of  $0.4\ \mathrm{mm}$. Fitting the interface to the two centrelines gives wall intersections at 18 and $17\ \mathrm{mm}$ \textcolor{black}{downstream} of the confluence, close to the observed \textcolor{black}{onset locations} of 17.6 and $16.8\ \mathrm{mm}$. The fitted interface rotates to $7^\circ$ past horizontal near $x\approx40\ \mathrm{mm}$, causing the upper band to dip below and the lower band to rise above mid-height, before settling at approximately $3.5\ \mathrm{mm}$ above the floor. The dye correlation predicts a similar distance, $L\approx22\ \mathrm{mm}$ at $Ri\approx0.9$, for the dyed stream to reach the opposite wall. Because the interface orientation is inferred from the \textcolor{black}{measured} band positions rather than computed independently, this agreement does not constitute proof of a \textcolor{black}{rotating} interface. However, it does show that the observed deposition pattern is consistent with such \textcolor{black}{a} mechanism. \Cref{sec:disc_mechanism} relates the fitted distances to the lock-exchange theory of Benjamin \cite{benjamin1968}.

\section{Discussion}
\label{sec:discussion}
\subsection{The simulated conditions were not at a small Richardson number}
\label{sec:disc_ri}

The simulations were set up on the assumption that a density difference of about 2\% is negligible. The criterion developed in \cref{sec:governing} can now be applied to the conditions that were actually simulated (\cref{tab:ri}). For the reference case, the simulated feed of $0.61\ \mathrm{mol\,kg_w^{-1}}$ closely \textcolor{black}{matches} the reacting pair used experimentally at $0.59\ \mathrm{mol\,kg^{-1}}$, with $\Delta\rho\approx20\ \mathrm{kg\,m^{-3}}$. At the simulated bulk velocity of $14.6\ \mathrm{mm\,s^{-1}}$ in the $8.5\ \mathrm{mm}$ channel, this density difference gives $U_b\approx41\ \mathrm{mm\,s^{-1}}$ and $Ri\approx7.8$. The buoyancy force \textcolor{black}{neglected} in the simulations is therefore not a \textcolor{black}{perturbation} to the reference case but a dominant dynamical effect. The same conclusion applies to the deposition experiments. Using the  estimated density difference of $11\ \mathrm{kg\,m^{-3}}$ between the $0.312\ \mathrm{M}$ feeds gives Richardson numbers of approximately 0.9, 3.4, and 14 across the experimental flow rate range. Consequently, \textcolor{black}{buoyancy} is expected to influence reactant transport in nearly all conditions, \textcolor{black}{except} for the density-matched experiments.

\begin{table}[htbp]
    \centering
    \caption{Density difference and Richardson number of the reacting conditions. $\Delta\rho$ of the $0.312\ \mathrm{M}$ feeds is interpolated between the measured reacting pairs of \cref{tab:density}; that of the simulated reference case is taken from the measured pair at $0.59\ \mathrm{mol\,kg^{-1}}$.}
    \label{tab:ri}
    \begin{tabular}{lcccccc}
        \toprule
        Condition & $H$ & $Q$ & $\Delta\rho$ & $U$ & $U_b$ & $Ri$ \\
        & $[\mathrm{mm}]$ & $[\mathrm{mL\,min^{-1}}]$ & $[\mathrm{kg\,m^{-3}}]$ & $[\mathrm{mm\,s^{-1}}]$ & $[\mathrm{mm\,s^{-1}}]$ & $[-]$ \\
        \midrule
        Simulation, reference case & 8.5 & 48.5 & $\approx20$ & 14.6 & 41 & 7.8 \\
        Deposition, S0 & 6.0 & 64.8 & $\approx11$ & 27.7 & 26 & 0.85 \\
                       & 6.0 & 32.3 & $\approx11$ & 13.8 & 26 & 3.4 \\
                       & 6.0 & 16.2 & $\approx11$ & 6.9 & 26 & 14 \\
        Density-matched, \textcolor{black}{S0} & \textcolor{black}{6.0} & 50 & $\approx0.1$ & \textcolor{black}{21.4} & \textcolor{black}{2.4} & \textcolor{black}{0.01} \\
        \bottomrule
    \end{tabular}
\end{table}

\textcolor{black}{The onset of the side-wall bands follows the order of these Richardson numbers.} At $Ri\approx3.4$ and 14, the buoyancy velocity exceeds the imposed velocity, and the bands appear from the confluence. At $Ri\approx0.85$, where the \textcolor{black}{buoyancy} and imposed velocities are of comparable magnitude, band formation is delayed \textcolor{black}{until} approximately $17\ \mathrm{mm}$ downstream. According to the dye correlation \cref{eq:collapse}, the distance required for the dense fluid to \textcolor{black}{reach the} opposite wall should differ by only a factor of about 2.5 between the two higher flow rate conditions, whereas the observed band onset distance changes by more than a factor of 20. The two measures, however, are not directly comparable. The dye \textcolor{black}{experiments} record the arrival of the dyed fluid at a wall in the projection, \textcolor{black}{whereas} the band onset marks the first appearance of solid deposition. In addition, the reacting feeds differ in composition from the dye solutions, so the \textcolor{black}{dye} correlation is not expected to predict the band \textcolor{black}{onset distance} quantitatively. The ordering \textcolor{black}{is nevertheless} consistent.

\subsection{Where the reactants meet the walls}
\label{sec:disc_mechanism}

The observations are consistent with a simple picture in which gypsum forms \textcolor{black}{where the interface between the two feeds meets} a wall (\cref{fig:fullcircle}). In the absence of buoyancy, the streams flow side by side and are separated by a vertical interface along the channel's centreline that meets the floor and the lid. That is where the constant-density simulations predict deposition (\cref{sec:cfd}), and where it appears in the density-matched experiment (\cref{sec:fullcircle}). When a  density difference is present, the denser feed underflows the \textcolor{black}{less dense one} and the interface rotates towards the horizontal, as observed for co-flowing streams in microchannels \cite{yoon2005}. Once the streams become vertically stratified, the interface intersects the two side walls near mid-height, where the bands are found, at $2.7$ to $3.6\ \mathrm{mm}$ in the $6\ \mathrm{mm}$ channel. At the highest flow rate, the band enters near the ceiling on one wall and near the floor on the other before converging towards mid-height over approximately $20\ \mathrm{mm}$, consistent with an interface that is still rotating downstream of the confluence. The earlier onset of the bands at larger $Ri$ is likewise consistent with faster buoyancy driven rotation of the interface. 

The observed stratification can be \textcolor{black}{interpreted} as a lock exchange turned on its side. Immediately downstream of the splitter plate, the two feeds lie side by side, analogous to the initial state of \textcolor{black}{a} lock-exchange experiment after removal of the gate. In this analogy, the channel height $H$ acts as the fluid depth and the half-width $W/2$ as the distance each front must travel. Benjamin \cite{benjamin1968} showed that the only steady gravity current between two horizontal boundaries that conserves energy occupies half the depth and advances at $c=\tfrac12\sqrt{g'H}$ relative to the fluid ahead. Such a current carries no head wave; the head that rises to about twice the mean thickness is \textcolor{black}{characteristic} of deeply submerged, dissipative gravity currents. If the front is advected downstream at the bulk velocity $U$, it reaches the opposite side wall after a distance $x\approx(W/2)\,U/c=W/\sqrt{Ri}$. For the experiment at $64.8\ \mathrm{mL\,min^{-1}}$ ($Ri\approx0.85$) this \textcolor{black}{estimate} gives $x\approx14\ \mathrm{mm}$, compared with  observed  band onsets of 17.6 and $16.8\ \mathrm{mm}$ (\cref{fig:fullcircle}b). Once the gravity current fronts reach the side walls, they are reflected and the interface can \textcolor{black}{overshoot} the horizontal before it relaxes towards its final position. The first interfacial seiche \textcolor{black}{(standing-wave)} mode of two equal layers spanning the channel width propagates at the same speed $c$ and has a period $2W/c\approx2.0\ \mathrm{s}$. Advected at the bulk velocity, this corresponds to a wavelength of $57\ \mathrm{mm}$. The fitted interface in \cref{fig:fullcircle}b has a wavelength of $74\ \mathrm{mm}$, which is \textcolor{black}{of} the same order. Consistent with this interpretation, the band on the wall reached by the Na$_2$SO$_4$ current rises to $4.3\ \mathrm{mm}$ before settling at $3.5\ \mathrm{mm}$, or $0.58\,H$, close to the half-depth predicted for Benjamin's energy-conserving current. The overshoot is modest, about one fifth of the final level rather than the factor of two increase characteristic of deep current heads, again suggesting a current occupying roughly half the channel depth. Both inviscid estimates underpredict the observed distances by approximately 20\,\%, which is plausible given viscous drag on the fronts in a $6\ \mathrm{mm}$ channel. In addition, the density difference for the $0.312\ \mathrm{M}$ feeds was interpolated rather than measured directly. This interpretation therefore remains tentative. It has not yet been tested in simulations that include buoyancy and the relative \textcolor{black}{contributions} of interface rotation and precipitation kinetics to the \textcolor{black}{asymmetry} between the two walls remain unresolved.

\subsection{Consequences for modelling}
\label{sec:disc_modelling}

The practical lesson is that a small difference in relative density is not, by itself, sufficient justification for neglecting buoyancy. Whether the buoyancy term can be ignored is determined by the Richardson number, which is proportional to the channel height and inversely proportional to the square of the velocity. At the flow rates at which \textcolor{black}{such flow cells} are typically operated, a density difference of a few tenths of a percent can already yield $Ri>1$. The \textcolor{black}{required} check is \textcolor{black}{straightforward}: it \textcolor{black}{needs} only the density difference, whether measured or estimated, together with the channel height and bulk velocity. \textcolor{black}{A quantitative version of this check follows from \cref{eq:collapse}: the dyed stream does not reach the opposite wall within a device of length $L_{\mathrm{dev}}$ if the predicted clear-zone length exceeds $L_{\mathrm{dev}}$, that is, if $Ri<(3.31\,H/L_{\mathrm{dev}})^{1/0.66}$. For the present cell ($L_{\mathrm{dev}}=120\ \mathrm{mm}$) this gives $Ri<0.065$, $0.11$ and $0.19$ for $H=6.0$, $8.5$ and $12.3\ \mathrm{mm}$, in agreement with the measured transition in \cref{fig:collapse}b. Because the correlation was obtained in one geometry, this threshold should be regarded as an estimate for similar confluence cells rather than a general criterion.} When the resulting Richardson number is \textcolor{black}{above this value}, \textcolor{black}{buoyancy} should be retained in the model. Under the conditions considered here, this adds little complexity. For $\Delta\rho/\rho_0\le2\,\%$,  the Boussinesq approximation remains valid, so the effect can be incorporated through a \textcolor{black}{buoyancy} term in \cref{eq:momentum}, \textcolor{black}{requiring} only one additional coefficient per species in the momentum equation.

For precipitation fouling, the implications are more consequential. A deposition model can act only on the reactants that the transport model delivers to the wall. If the transport model predicts the interface between the reactant streams on the wrong wall, no adjustment of the deposition kinetics, whether through a rate constant, supersaturation threshold, or other fitting parameter, can recover the observed deposition pattern. Any parameter values obtained by fitting such a model would therefore lack physical meaning. The location of a deposit thus provides a direct and stringent test of the transport description. Before a deposition law is calibrated, the model must first demonstrate that it predicts where the reactants reach the wall \textcolor{black}{and} where deposition occurs. Only once this transport behaviour is captured can the deposition kinetics be interpreted with confidence. 
\section{Conclusion}
\label{sec:conclusion}
We have tested the common assumption that small density differences can be neglected when modeling the mixing of reacting streams, using a \textcolor{black}{confluence flow cell} to study gypsum fouling. A constant-density simulation of the cell predicts the deposit on the floor and lid along the channel centreline, whereas the experiments show gypsum bands on the side walls \textcolor{black}{that begin closer to the confluence} as the flow rate decreases. Dye experiments with density differences of 0.14 to 2.1\,\% demonstrate that buoyancy transports the dense stream across the channel: the front velocity depends strongly on the density difference and not on the flow rate, while the distance at which the dyed stream reaches the opposite wall collapses on the bulk Richardson number according to $L/H=3.31\,Ri^{-0.66}$ over four decades of $Ri$. When the stream densities match, this cross-channel transport is suppressed, and in a density-matched reacting experiment gypsum deposits on the floor, consistent with the constant-density simulation prediction.

The simulated reference case itself lies at $Ri\approx8$, while the deposition experiments span $Ri\approx1$ to 14. The condition required to justify the constant-density approximation was therefore not satisfied, despite a relative density difference being only about 2 \%. A small relative density difference is necessary but not sufficient for neglecting buoyancy; the Richardson number must also be small. Because evaluating $Ri$ requires only the density difference, channel height, and flow velocity, \textcolor{black}{it is recommended to} perform this check before applying a constant-density model to mixing, reaction, or fouling problems.


\small
\nomenclature[A01]{$A$}{Molality of dissolved Ca$^{2+}$ (scalar field) [$\mathrm{mol\,kg_w^{-1}}$]}
\nomenclature[A02]{$A_f$}{Area of wall face $f$ [$\mathrm{m^2}$]}
\nomenclature[A03]{$B$}{Molality of dissolved SO$_4^{2-}$ (scalar field) [$\mathrm{mol\,kg_w^{-1}}$]}
\nomenclature[A04]{$C$}{Molality of precipitated CaSO$_4$ (bookkeeping field) [$\mathrm{mol\,kg_w^{-1}}$]}
\nomenclature[A04b]{$c$}{Front speed of a half-depth gravity current, $\tfrac12\sqrt{g'H}$ [$\mathrm{m\,s^{-1}}$]}
\nomenclature[A05]{$C_{\mathrm{mix}}$}{Equimolar concentration giving the target saturation ratio [$\mathrm{mol\,kg_w^{-1}}$]}
\nomenclature[A06]{$C_1,\ C_2,\ C_\mu$}{Coefficients of the $k$--$\varepsilon$ model}
\nomenclature[A07]{$D$}{Hydraulic diameter of a wide inlet channel, $D=2h$ [$\mathrm{m}$]}
\nomenclature[A08]{$D_A,\ D_B$}{Molecular diffusivities of Ca$^{2+}$ and SO$_4^{2-}$ [$\mathrm{m^2\,s^{-1}}$]}
\nomenclature[A09]{$E$}{Expansion ratio, $E=H/h=1+S/h$}
\nomenclature[A10]{$F$}{Dye signal, $\log[R/\tfrac12(G+B)]$ of the colour channels $R$, $G$, $B$}
\nomenclature[A11]{$f_A,\ f_B$}{Area fractions of inlets A and B}
\nomenclature[A12]{$\mathbf{g},\ g$}{Gravitational acceleration [$\mathrm{m\,s^{-2}}$]}
\nomenclature[A13]{$g'$}{Reduced gravity, $g\,\Delta\rho/\rho_0$ [$\mathrm{m\,s^{-2}}$]}
\nomenclature[A14]{$H$}{Channel height downstream of the step [$\mathrm{m}$]}
\nomenclature[A15]{$h$}{Height of the upstream channel [$\mathrm{m}$]}
\nomenclature[A16]{$h_{\mathrm{dep}}$}{Equivalent gypsum deposit thickness [$\mathrm{m}$]}
\nomenclature[A17]{$j$}{Wall deposition flux, limited by diffusive supply [$\mathrm{mol\,m^{-2}\,s^{-1}}$]}
\nomenclature[A18]{$j_{\mathrm{kin}}$}{Kinetic (intrinsic) wall deposition flux [$\mathrm{mol\,m^{-2}\,s^{-1}}$]}
\nomenclature[A19]{$k$}{Turbulent kinetic energy [$\mathrm{m^2\,s^{-2}}$]}
\nomenclature[A20]{$k_{\mathrm{bulk}}$}{Rate constant of bulk precipitation [$\mathrm{mol\,kg_w^{-1}\,s^{-1}}$]}
\nomenclature[A21]{$k_{\mathrm{wall}}$}{Rate constant of wall deposition [$\mathrm{mol\,m^{-2}\,s^{-1}}$]}
\nomenclature[A22]{$L$}{Clear-zone length [$\mathrm{m}$]}
\nomenclature[A23]{$L_c$}{Length of the channel insert [$\mathrm{m}$]}
\nomenclature[A24]{$M_{\mathrm{gyp}}$}{Molar mass of gypsum [$\mathrm{kg\,mol^{-1}}$]}
\nomenclature[A25]{$N_{\mathrm{dep}}$}{Cumulative deposited amount per unit wall area [$\mathrm{mol\,m^{-2}}$]}
\nomenclature[A26]{$n_{\mathrm{bulk}},\ n_{\mathrm{wall}}$}{Reaction orders of bulk precipitation and wall deposition}
\nomenclature[A27]{$\Delta n_f$}{Amount deposited on face $f$ in one time step [$\mathrm{mol}$]}
\nomenclature[A28]{$P$}{Pressure relative to hydrostatic [$\mathrm{Pa}$]}
\nomenclature[A29]{$p$}{Kinematic pressure, $P/\rho_0$ [$\mathrm{m^2\,s^{-2}}$]}
\nomenclature[A30]{$P_k$}{Production of turbulent kinetic energy [$\mathrm{m^2\,s^{-3}}$]}
\nomenclature[A31]{$Q$}{Volumetric flow rate per inlet [$\mathrm{mL\,min^{-1}}$]}
\nomenclature[A32]{$R^2$}{Coefficient of determination}
\nomenclature[A33]{$\mathrm{Re}_D$}{Reynolds number of the inlet channel, $U_{\mathrm{in}}D/\nu$}
\nomenclature[A34]{$\mathrm{Re}_H$}{Reynolds number downstream of the step, $UH/\nu$}
\nomenclature[A35]{$Ri$}{Bulk Richardson number, $g'H/U^2$}
\nomenclature[A37]{$r_{\mathrm{bulk}}$}{Rate of precipitation in the liquid [$\mathrm{mol\,kg_w^{-1}\,s^{-1}}$]}
\nomenclature[A38]{$r_{\mathrm{wall}}$}{Rate of deposition in wall-adjacent cells [$\mathrm{mol\,kg_w^{-1}\,s^{-1}}$]}
\nomenclature[A39]{$S$}{Step height [$\mathrm{m}$]}
\nomenclature[A40]{$S_A,\ S_B,\ S_C$}{Source terms of the scalar fields $A$, $B$, $C$ [$\mathrm{mol\,kg_w^{-1}\,s^{-1}}$]}
\nomenclature[A41]{$\mathrm{SR}$}{Saturation ratio}
\nomenclature[A42]{$t,\ \Delta t$}{Time, time step [$\mathrm{s}$]}
\nomenclature[A43]{$\mathbf{U}$}{Mean velocity vector [$\mathrm{m\,s^{-1}}$]}
\nomenclature[A44]{$U$}{Bulk velocity downstream of the step, $2Q/(WH)$ [$\mathrm{m\,s^{-1}}$]}
\nomenclature[A45]{$U_{\mathrm{in}}$}{Bulk velocity in the upstream channel [$\mathrm{m\,s^{-1}}$]}
\nomenclature[A46]{$U_b$}{Buoyancy velocity, $\sqrt{g'H}$ [$\mathrm{m\,s^{-1}}$]}
\nomenclature[A48]{$V_c$}{Volume of a wall-adjacent cell [$\mathrm{m^3}$]}
\nomenclature[A49]{$v_{\mathrm{app}}$}{Apparent speed of the dye front [$\mathrm{m\,s^{-1}}$]}
\nomenclature[A50]{$W$}{Channel width [$\mathrm{m}$]}
\nomenclature[A51]{$X_{\mathrm{crit}}$}{Critical saturation ratio}
\nomenclature[A52]{$x,\ y,\ z$}{Streamwise, spanwise and vertical coordinates [$\mathrm{m}$]}
\nomenclature[A53]{$x_r$}{Reattachment length [$\mathrm{m}$]}
\nomenclature[A54]{$x_{\mathrm{step}}$}{Streamwise position of the step [$\mathrm{m}$]}

\nomenclature[B01]{$\delta_f$}{Reciprocal wall-normal cell distance of face $f$ [$\mathrm{m^{-1}}$]}
\nomenclature[B02]{$\Delta\rho$}{Density difference between the two streams [$\mathrm{kg\,m^{-3}}$]}
\nomenclature[B03]{$\varepsilon$}{Dissipation rate of turbulent kinetic energy [$\mathrm{m^2\,s^{-3}}$]}
\nomenclature[B04]{$\nu$}{Kinematic viscosity [$\mathrm{m^2\,s^{-1}}$]}
\nomenclature[B05]{$\nu_t$}{Eddy viscosity [$\mathrm{m^2\,s^{-1}}$]}
\nomenclature[B06]{$\rho$}{Density [$\mathrm{kg\,m^{-3}}$]}
\nomenclature[B07]{$\rho_0$}{Reference density [$\mathrm{kg\,m^{-3}}$]}
\nomenclature[B08]{$\rho_{\mathrm{gyp}}$}{Density of gypsum [$\mathrm{kg\,m^{-3}}$]}
\nomenclature[B09]{$\rho_w$}{Density of water [$\mathrm{kg\,m^{-3}}$]}
\nomenclature[B11]{$\sigma,\ \sigma_w$}{Relative supersaturation in the bulk and at the wall}
\nomenclature[B12]{$\sigma_k,\ \sigma_\varepsilon$}{Turbulent Prandtl numbers for $k$ and $\varepsilon$}

\nomenclature[C01]{$0$}{Reference state}
\nomenclature[C02]{$A,\ B$}{Species or inlet A, B}
\nomenclature[C03]{$\mathrm{app}$}{Apparent}
\nomenclature[C04]{$\mathrm{bulk}$}{In the liquid}
\nomenclature[C05]{$c$}{Computational cell}
\nomenclature[C06]{$\mathrm{crit}$}{Critical}
\nomenclature[C07]{$\mathrm{dep}$}{Deposit}
\nomenclature[C08]{$f$}{Wall face}
\nomenclature[C09]{$\mathrm{gyp}$}{Gypsum}
\nomenclature[C10]{$\mathrm{in}$}{Upstream (inlet) channel}
\nomenclature[C11]{$\mathrm{kin}$}{Kinetic}
\nomenclature[C12]{$\mathrm{mix}$}{Mixed}
\nomenclature[C13]{$n,\ n+1$}{Time levels}
\nomenclature[C14]{$r$}{Reattachment}
\nomenclature[C15]{$\mathrm{step}$}{Backward-facing step}
\nomenclature[C16]{$t$}{Turbulent}
\nomenclature[C17]{$w$}{Water ($\rho_w$, $\mathrm{kg_w}$) or wall ($\sigma_w$)}
\nomenclature[C18]{$\mathrm{wall}$}{At the wall}
\nomenclature[C19]{$T$}{Transpose}

\nomenclature[D01]{DDES-SA}{Delayed detached-eddy simulation based on the Spalart--Allmaras model}
\nomenclature[D02]{GAMG}{Geometric--algebraic multigrid solver}
\nomenclature[D03]{LES}{Large-eddy simulation}
\nomenclature[D04]{micro-PIV}{Micro particle image velocimetry}
\nomenclature[D05]{PIMPLE}{Merged PISO--SIMPLE pressure--velocity algorithm}
\nomenclature[D06]{RANS}{Reynolds-averaged Navier--Stokes}
\nomenclature[D07]{S0, S2.5, S6.3}{Inserts with step heights $S=0$, 2.5 and 6.3 mm}
\nomenclature[D08]{SR}{Saturation ratio}
\nomenclature[D09]{WALE}{Wall-adapting local eddy-viscosity subgrid model}
\nomenclature[D10]{$k$--$\varepsilon$}{Standard two-equation turbulence model}

\printnomenclature

\section*{\textcolor{black}{CRediT authorship contribution statement}}
\textcolor{black}{\textbf{Isaac Appelquist L{\o}ge:} Conceptualization, Methodology, Software, Formal analysis, Investigation, Data curation, Visualization, Writing -- original draft, Funding acquisition. \textbf{Candela Almaz\'an P\'erez:} Investigation, Writing -- review \& editing. \textbf{Maike W. Baltussen:} Methodology, Formal analysis, Supervision, Writing -- review \& editing. \textbf{Nadia Shardt:} Formal analysis, Supervision, Writing -- review \& editing, Funding acquisition.}

\section*{Data Availability}
The data that support the findings of this study are available from the corresponding author upon reasonable request. 

\section*{Acknowledgements}
The work was supported by an \href{https://www.carlsbergfondet.dk/det-har-vi-stoettet/cf24-1619/}{Internationalisation Grant} from the Carlsberg Foundation under grant number CF24-1619. The Department of Chemical Engineering at NTNU most generously hosted this work. \textcolor{black}{The authors thank Sebastien Charles Roger Simon and Jens Norman for assistance in the laboratory, and for discussing both troubleshooting and of the results.}

\bibliographystyle{Set-up/elsarticle-num_v1}
\bibliography{Set-up/References}

\end{document}